\documentclass{aa}
\usepackage{graphicx}
\usepackage{natbib}
\usepackage{color}
\usepackage{float}
\usepackage{txfonts}
\usepackage{lscape}

\begin{document}

\title{Multiply eclipsing candidates from the TESS satellite}

\author{Zasche,~P.~\inst{1},
        Henzl,~Z.~\inst{2,3},
        Ma\v{s}ek,~M.~\inst{2,4}}

\offprints{Petr Zasche, \email{zasche@sirrah.troja.mff.cuni.cz}}

 \institute{
  $^{1}$ Astronomical Institute, Charles University, Faculty of Mathematics and Physics, V~Hole\v{s}ovi\v{c}k\'ach 2, CZ-180~00, Praha 8, Czech Republic\\
  $^{2}$ Variable Star and Exoplanet Section, Czech Astronomical Society, Fri\v{c}ova 298, CZ-251 65 Ond\v{r}ejov, Czech Republic\\
  $^{3}$ Hv\v{e}zd\'arna Jaroslava Trnky ve Slan\'em, Nosa\v{c}ick\'a 1713, CZ-274 01 Slan\'y 1, Czech Republic\\
  $^{4}$ FZU - Institute of Physics of the Czech Academy of Sciences, Na Slovance 1999/2, CZ-182~21, Praha, Czech Republic
 }

\titlerunning{Catalogue of TESS multiple eclipsers}
\authorrunning{Zasche \& Henzl \& Ma\v{s}ek}

  \date{\today}

\abstract{ We present the catalogue of the TESS targets showing multiple eclipses. It means that in all
of these stars we detected two sets of eclipses, for which their two distinctive periods can be
derived. These multiple stellar systems can be either doubly eclipsing quadruples, or triple-star
coplanar systems showing besides the inner eclipses also the eclipses on the outer orbit. In total, 116
systems were found as doubly eclipsing, while 25 stars were identified as triply eclipsing triples.
Several confirmed blends of two close sources were not included into our analysis. All these systems
were identified scanning the known eclipsing systems taken from VSX database checking their TESS light
curves. The average period of the dominant pair A is 2.7 days in our sample, while for the second pair
B the average period is 5.3 days. Several systems show evident ETV changes even from the short interval
of the TESS data, indicating possible period changes and short mutual orbit. We also present an
evidence that the system V0871 Cen is probably a septuple-star system of architecture (Aa-Ab)-B-C-D.
Most of the presented systems are adequately bright and showing deep enough eclipses, hence we call for
new ground-based observations for these extremely interesting multiples. Owing to this motivation our
catalog contains besides the ephemerides for both pairs also their depths of eclipses and the light
curve shapes as extracted from the TESS data. These new ground based observations would be very useful
for further derivation of the mutual movement of both pairs on their orbit via detection of the ETVs of
both pairs for example.}


\maketitle

\section{Introduction} \label{intro}

It is more than 200 years ago, when there was proposed a hypothesis that the true origin of brightness
changes of Algol is caused by an obscuring body and its orbit around the main star
\citep{1783RSPT...73..474G}. This was a moment when the eclipsing binary research was born. However, it
lasted about 2 centuries more when become obvious that the use of eclipsing binaries bring us
absolutely unique insight into the basic stellar properties like masses, radii, luminosities, etc.
Their role in current astrophysical research is undisputable (see e.g. \citealt{2012ocpd.conf...51S}).
And nowadays in the era of huge photometric surveys and projects their number is increasing rapidly,
hence can be used for calibrating already existing stellar models of evolution.

Besides that, thanks to large surveys, due to significantly reducing the scatter, and tremendously
increasing the number of known systems, we are also able to discover and study much more complicated
objects. For example the huge survey OGLE \citep{1992AcA....42..253U} was used to identify about a
thousand new candidates on triple stars due to their study of period changes via a so-called
eclipse-timing variation (ETV) method, see \cite{2019MNRAS.485.2562H}.

Moreover, quite a new group of objects nowadays known as doubly-eclipsing ones were also mostly being
discovered thanks to large photometric surveys like OGLE, Kepler, Corot, or TESS (see e.g.
\citealt{2016AcA....66..405S}, \citealt{2012A&A...541A.105L}, \citealt{2017MNRAS.471.1230H}, and
\citealt{2021ApJ...917...93K}). These objects show us two distinct eclipsing periods coming from one
point source on the sky. The study of such stars should bring us fresh insight to the stellar formation
mechanisms (see e.g. \citealt{2021Univ....7..352T}). Nevertheless, if both these eclipsing binaries are
really bound to each other and share a common orbit around their barycenter, both should also share the
same distance, same age, and same metallicity, obviously. This is an important aspect, which should be
tested when analysing a particular system, but also set tight constraints to our model.

Our main motivation for study and discovery of such systems is the fact that they represent ideal
astrophysical laboratories. Thanks to them we can study the celestial mechanics in "real time", the
dynamical influence between the inner and outer orbits, Kozai cycles, dynamical evolution of the orbits
like its precession and inclination changes, etc. Moreover, when having enough such 2+2 quadruples with
known orbits, one can study their origin, and subsequent evolution. Are they products of disk
fragmentation, or an N-body dynamics? And what about the suspected mean motion resonances of both inner
pairs? Many open questions waiting to be answered, and only much more robust sample of them can bring
us some clue.

\section{The selection process}

The first doubly eclipsing system confirmed showing two eclipsing periods was V994 Her
\citep{2008MNRAS.389.1630L}. Nowadays the group comprises about 160 doubly eclipsing systems with both
periods known. However, for not all of them there was proved that these are not only blends of two
close-by components not connected gravitationally. Those ones, where also their mutual orbit is known
are still only a few. Especially in dense star fields this is a serious problem: one cannot definitely
prove that the signal of two periods really come to our telescope from one point source on the sky.
Several dozens of doubly eclipsing systems were discovered in dense LMC and SMC fields, as well as in
closer Galactic bulge fields by the large survey OGLE (\citealt{2011AcA....61..103G},
\citealt{2013AcA....63..323P}, and \citealt{2016AcA....66..405S}).

Besides that, quite recently there was discovered that quite a significant fraction of new doubly
eclipsing binaries are found in already known and bright eclipsing binaries. The problem usually is
that these additional eclipses were missed because of their small amplitude (compared with the dominant
pair, like in V482 Per, \citealt{2017ApJ...846..115T}), or due to the fact that these stars are
sometimes easily too bright for nowadays telescopes (like BU CMi, \citealt{2021tsc2.confE..14J}). Maybe
the most remarkable example is the system BG Ind, studied in detail with ground-based photometry
\citep{2011MNRAS.414.2479R} having not noticed anything suspicious. But using the TESS data it was
discovered to be the nearest doubly eclipsing system \citep{2021MNRAS.503.3759B}. It was being missed
simply due to too shallow photometric amplitude of pair B. Sometimes only small parts of the light
curves near eclipses are being observed, and the rest is being monitored only very rarely, which led to
undetection of the additional eclipses (like for V498 Cyg, see \citealt{2022arXiv220102516S}).

Taking into account all these aspects and limitations, we decided to use the super-precise TESS data
and to scan as much potential systems as possible. Knowing the fact that nowadays about a million
eclipsing binaries is already known \citep{2006SASS...25...47W}, we decided to use the super-precise
and freely available database of the TESS satellite \citep{2015JATIS...1a4003R}. Besides that, probably
the most complete database of information on variable stars running under AAVSO, named VSX
\citep{2006SASS...25...47W} was used. Our method was quite simple: scanning all known eclipsing
binaries found in the VSX with known periods, and trying to identify some additional eclipses in the
TESS data. Only those ones with given magnitudes brighter than 15 mag were considered. All EA-type
stars from the VSX have been checked for additional eclipses. Such a group of stars comprise huge
majority of systems in our sample. Among the others there were identified only several other stars with
EB or EW classification with periods longer than 0.5 days. Besides that, also all systems from the CzeV
catalogue \citep{2017OEJV..185....1S} were plotted and checked.

For some of the systems the additional eclipses were already seen directly when plotting TESS fluxes
versus time, for some others we needed first to subtract the dominant eclipsing binary shape and later
to search for additional eclipses on the residuals. Plotting the TESS photometric data in the phased
light curve with particular period is an easy task with the available program  named {\tt{lightkurve}}
\citep{2018ascl.soft12013L}. An eye-detection technique was found to be the most effective for finding
these additional eclipses in the data. A human eye is a very sensitive tool when looking for some
additional patterns in the already very periodic signal like the ordinary eclipsing binaries are. We
are aware of its limitations, as well as of limitation of the large TESS pixels (see below for more
detailed discussion).

\section{Method}

In total, about 70000 stars were checked. Their data downloaded, their light curves plotted, and
checked for additional eclipses. Many suspected cases were studied in more detail, trying to definitely
confirm whether the strange behaviour of the TESS data is due to improper reduction, or real brightness
changing effect. Results of our in-deep scanning of the VSX/TESS are being summarized in the next
section.

Our goal was not the detailed analysis of these systems, but only to present a catalogue of candidate
systems. I.e. such stars, where with one simple eclipsing period one cannot easily describe all the
eclipses found in the TESS data. In total: 141 such stars were found, among them 116 doubly eclipsing
ones, and 25 candidates of triply eclipsing triple stars. Several other systems were also found as
suspected, but still rather questionable of its nature (most often the contact W UMa-type shape and
short periodic pulsation patterns are too similar).

Or method of disentangling the complete photometry into two separate light curves of both pairs was
rather straightforward. We used the program named Silicups (version
2.99\footnote{https://www.gxccd.com/cat?id=187\&lang=405}), which uses a phenomenological model for the
phased light curve. Such a preliminary light curve fit can be subtracted and one easily see, whether
some additional eclipses are also visible on the residual light curve. If so, we used a period
searching algorithm to detect the secondary period $P_B$ (using a PDM method).

 \begin{figure}
  \centering
  \includegraphics[width=0.46\textwidth]{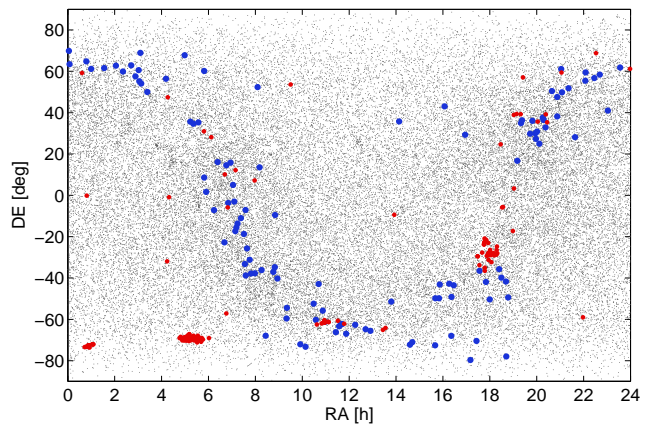}
  \caption{Sky distribution of our candidate stars (blue) together with all known doubly eclipsing systems nowadays (red), plotted
  together with other stars from TESS (only 50000 brightest ones).}
  \label{RADEsky}
 \end{figure}

As one can see from our Fig. \ref{RADEsky}, our detected candidate doubly eclipsing systems are of high
importance. Up to now, most of the detected doubly eclipsing systems were found in the OGLE fields, in
both Magellanic clouds LMC \& SMC, and the Galactic Dics and Bulge. However, these OGLE data provide
somehow biased photometric data (for example observing strategy, data cadency), hence also the detected
doubly eclipsing systems are definitely slightly different than that ones from TESS data (almost
complete detection of binaries with P $<$ 27 days). Two aspects are the most important here. At first,
the data cadency of the OGLE data are typically one observation per night only, but the overall time
span (of LMC \& SMC) is more than 10 years. On the other hand, the TESS data provide continuous
undisturbed photometry during 26 days. And the second important aspect is obviously the typical
scatter, or error of individual data point. OGLE provides data with about 0.01 mag scatter, while TESS
definitely order of magnitude better. Therefore, also the level of eclipse depth detectable in both
surveys should be rather different and many of the systems detected by our method here definitely
cannot be visible in the OGLE data. And finally, also the angular resolution of the TESS and the OGLE
data is very different.

On the other hand, what Fig. \ref{RADEsky} also shows us is the fact that our sample of newly detected
systems (blue ones in the figure) show much higher tendency of being closer to the Galactic disc than
the rest of other bright TESS targets. Qualitatively, the stars closer than 10$^\circ$ from the disc
comprise about 3/4 within our sample, however only 3 times lower about 1/4 of the TESS stars. It is
therefore questionable whether the number of doubly eclipsing binaries is higher in the Galactic disc
(due to higher star density, and hence higher probability of the blending?). However, we prefer the
other explanation namely the fact that all the stars in our sample comes from the already known
eclipsing systems, which are mostly being discovered in large photometric surveys detecting most of its
variables preferably closer to the Galactic disc.

Concerning the two periods, typically the more prominent one is for the pair A (usually having deeper
eclipses), but some exceptions exist. Our primary period of pair A is always the one as given in the
VSX catalog, and the pair B's period is the new derived one.

One can ask how we deal with the blending problem within the large TESS pixels. In huge majority of the
systems we were also able to identify both sets of eclipses on other ground based older photometry with
better angular resolution. These were namely the ASASsn (\citealt{2017PASP..129j4502K},
\citealt{2014ApJ...788...48S}), ZTF \citep{2019PASP..131a8003M}, or SWASP \citep{2006PASP..118.1407P}
data. Especially the ZTF survey was very useful source of photometry due to its high angular resolution
helping us to rule out the blending of the two close-by sources. Several blends of two close stars were
also proved as a by-product of our analysis (like V0432 CMa, ASASSN-V J020003.56+452605.2, and ASASSN-V
J123052.13-475634.5).

\begin{figure}
  \centering
  \includegraphics[width=0.32\textwidth]{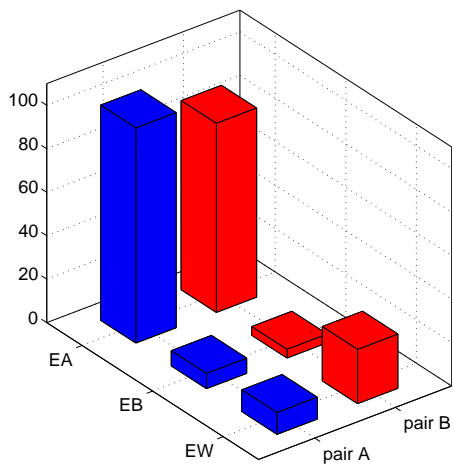}
  \caption{The distribution of individual eclipsing binary types from our detected sample. As one can
  see, huge majority are the EA+EA systems, which are being most easily discovered in the TESS data.}
  \label{3Dplot}
 \end{figure}

\section{Results}

In total we detected 116 systems as candidate 2+2 quadruple doubly eclipsing ones, huge majority of
them should be characterized as EA + EA (see Table \ref{systemsInfoDEBs}, and Fig. \ref{3Dplot}). This
phenomenological classification only tells us that our method is preferably able to discover such
systems more easily - just because the two periodic curves are being more easily disentangled into the
two periods and their shape is definitely the "eclipse-like". All of the disentangled light curves of
both pairs are being plotted in Figs. \ref{LC01} to \ref{LC08}. For all of the systems we left with its
original name within the VSX database and gave also their TIC numbers for identification and
coordinates in Table \ref{systemsInfoDEBs}.

\begin{figure}
  \begin{picture}(280,150)
  \put(-3,0){
  \includegraphics[width=0.25\textwidth]{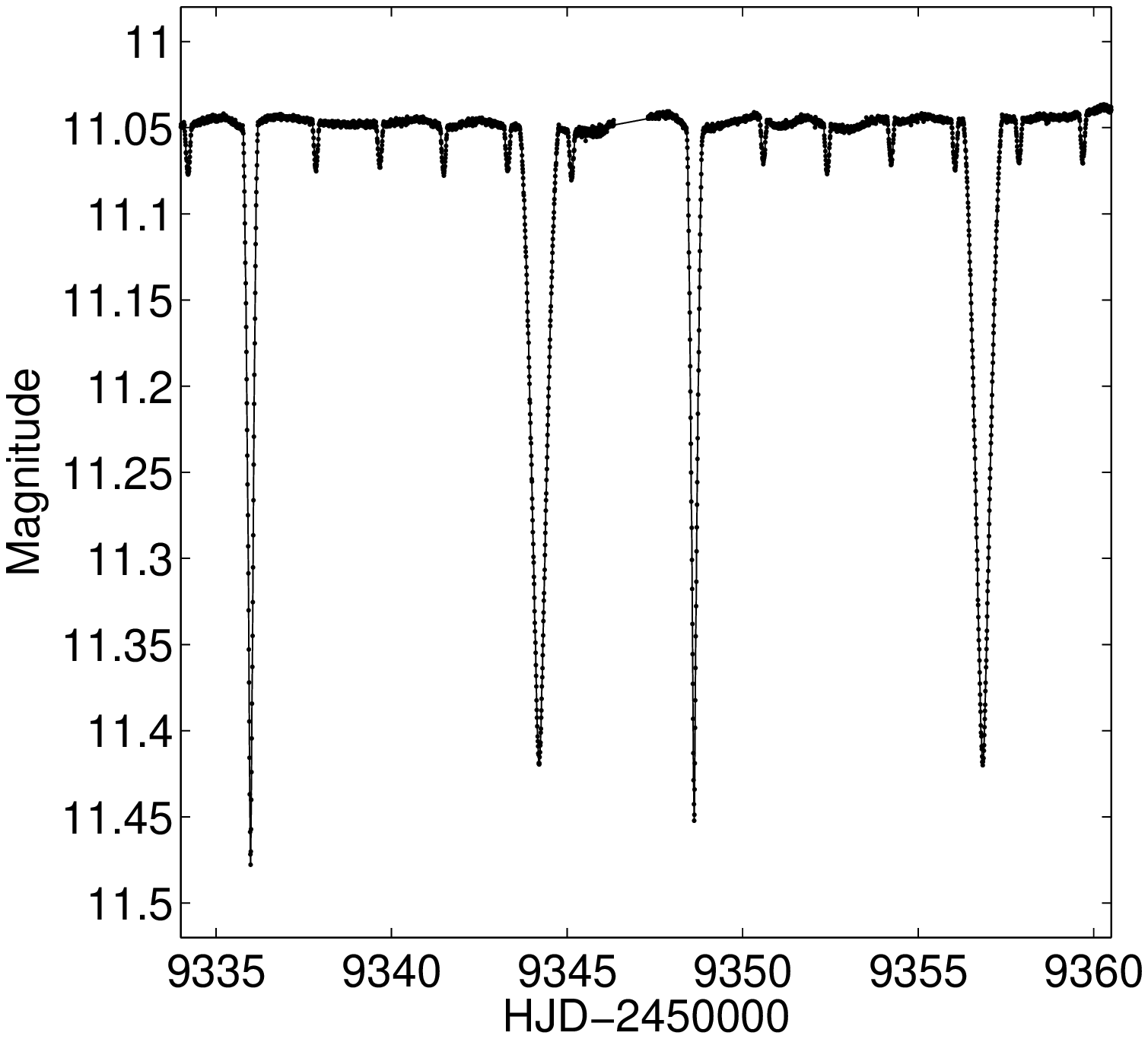}}
  \put(126,0){
  \includegraphics[width=0.25\textwidth]{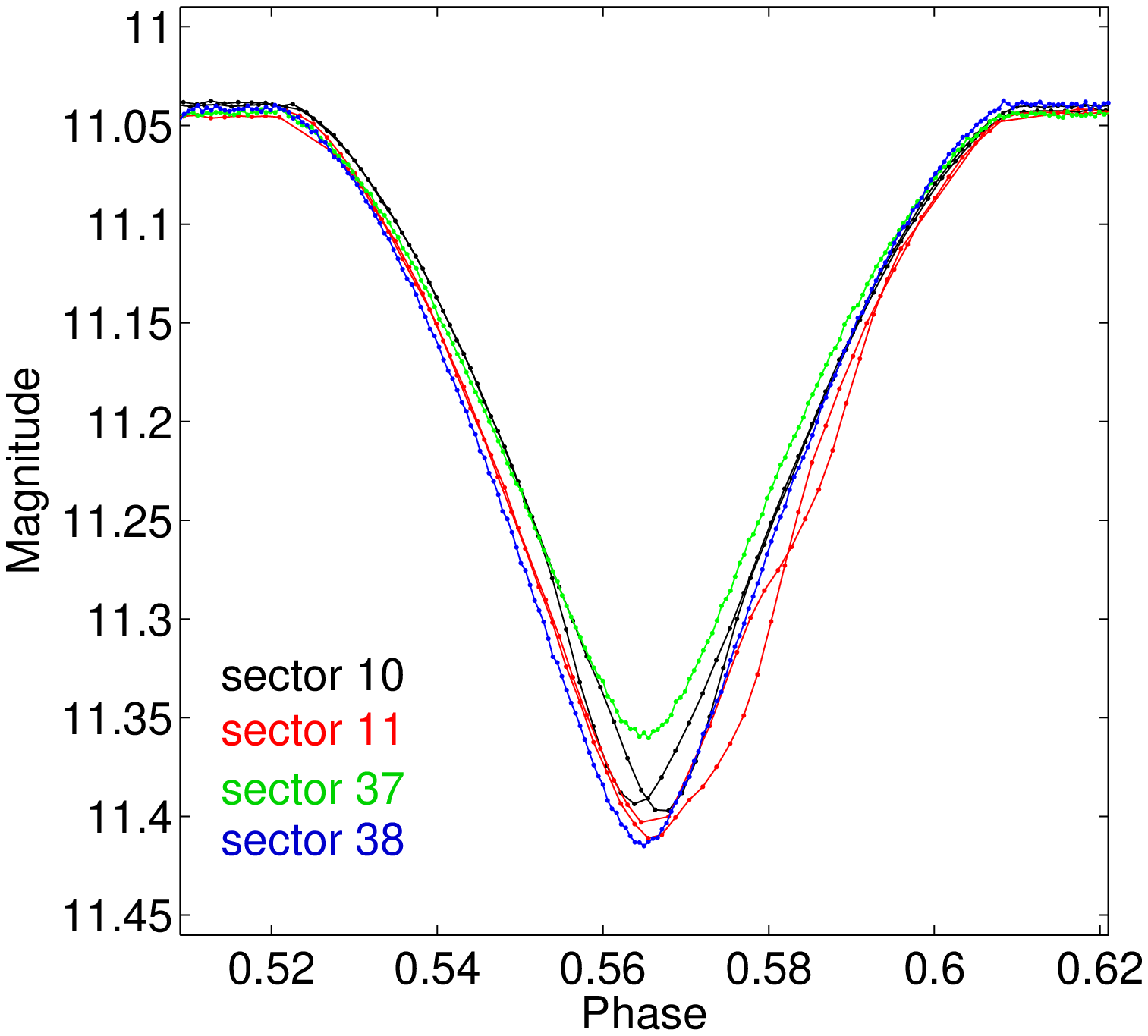}}
  \end{picture}
  \caption{Sample light curve of V0384 Cen from the TESS data. Left-hand side figure shows data from sector 38 with eclipses of both pairs in time. Right figure is phased only part of the curve near primary eclipses of pair A, showing how the pair B distorts the shape of the eclipse in different sectors of data. Moreover, also the depth of eclipse is obviously different in different sectors. This can be caused by two different phenomena. At first, the apsidal motion and change of omega on the eccentric orbit, and secondly an artifact from the TESS reduction causing different third light values in different sectors of data.}
  \label{FigLC_V384Cen}
 \end{figure}

In the Figure \ref{FigLC_V384Cen} we show for illustration how difficult sometimes should be to detect
the additional eclipses of pair B from the ground when having only poorly sampled data. This is quite
known eccentric system V0384 Cen, observed many times in the past, but nobody has noticed so far that
the shape of the minima are being distorted and such deviations are periodic coming from the unseen
additional pair. This is a typical example of bright systems observed in the past, but where the other
pair has order of magnitude smaller photometric amplitude. Unfortunately, the observers usually observe
only the eclipses, or even its bottom parts. In such cases like this one, the detection of the pair B
should be quite difficult. And here comes the big advantage of the continuous TESS data observing the
target with its superb precision for many days in the row.

The systems classified as triply eclipsing triples (see e.g. quite recent study by
\citealt{2022MNRAS.510.1352B}) were distinguished from the rest of other doubly eclipsing quadruples
easily due to their shapes of the additional eclipses (a "double peak") and also their uniqueness (i.e.
long orbits, long periods, only rarely observed in the TESS data). All of them are given below in Table
\ref{systemsInfoTriples}. The other ground based photometric surveys were usually also used for
detecting the third body outer periods. This was partly successful for several systems (due to their
large depths and long-lasting monitoring in these surveys) and we give their outer periods also in the
last column of Table \ref{systemsInfoTriples}. Their light curves showing besides the "ordinary"
eclipses also the additional ones are given in Fig. \ref{LCtroj}.

Probably the most interesting system seems to be ASASSN-V J101237.44-594344.8. It shows large
variations of ETV for the inner pair A and also exhibits eclipse depth variations (both in the TESS
data as well as in the older ground-based photometry).

In the following list we briefly mention those systems, which were found to be interesting in some
aspect as resulted from our analysis. Some of them were found to be located really close to the mean
motion resonance configuration, as already proposed in our previous paper \citep{2019A&A...630A.128Z},
and studied also theoretically by \citep{2020MNRAS.493.5583T}.

\begin{itemize}
    \item[$\bullet$] ASASSN-V J004727.28+644904.9 : very eccentric pair B (secondary eclipse in phase 0.705)
    \item[$\bullet$] ASASSN-V J020306.68+624315.4 : very eccentric pair B (secondary eclipse in phase 0.26)
    \item[$\bullet$] ASASSN-V J024221.82+625403.6 : almost exactly resonance 2:1 (only 0.2\% off)
    \item[$\bullet$] V1018 Cas : eccentric pair A, close to 4:3 resonance
    \item[$\bullet$] V0417 Aur : change of classification here. Former classification as pulsating-eclipsing oEA is now more probable as EA + EW system
    \item[$\bullet$] ASASSN-V J064048.28-224659.0 : eclipses of pair B are visible in TESS data only in 2020, missing in 2019, and 2018 (orbital precession?)
    \item[$\bullet$] ASASSN-V J071131.63-153341.3 : eccentric pair A, close to 2:3 resonance
    \item[$\bullet$] ASASSN-V J072304.90-110043.5 : eccentric pair B
    \item[$\bullet$] V0674 Pup : eccentric pair B
    \item[$\bullet$] WISE J075848.7-374315 : possible period changes
    \item[$\bullet$] CPD-34 3002 : slightly eccentric pair A
    \item[$\bullet$] ASASSN-V J091951.17-593306.9 : eccentric system (secondary eclipse in phase 0.44)
    \item[$\bullet$] ASASSN-V J092031.34-542438.1 : eccentric pair B
    \item[$\bullet$] WISE J100820.0-731554 : period changes; eccentric pair B, close double star according to GAIA (probably bound - similar $\pi$, and PM) 
    \item[$\bullet$] ASAS J103449-6013.1 : double star, 2 components only about 2.3$^{\prime\prime}$ distant, eccentric pair A
    \item[$\bullet$] ASASSN-V J105824.33-611347.6 : very fast apsidal motion or other ETV phenomena for pair A, eccentric also pair B
    \item[$\bullet$] ASAS J113426-6320.0 : slightly eccentric pair A, very eccentric pair B (sec min in phase 0.72)
    \item[$\bullet$] V0384 Cen : well-known eccentric binary A, see Fig. \ref{FigLC_V384Cen}
    \item[$\bullet$] KELT KS38C016096 : asymmetric light curve shape of pair A (spots?)
    \item[$\bullet$] ASASSN-V J125427.31-653437.7 : possible period changes
    \item[$\bullet$] ASASSN-V J155157.55-430547.7 : eccentric pair B
    \item[$\bullet$] ASASSN-V J173344.14-363037.8 : possible period changes, eccentric B
    \item[$\bullet$] ASASSN-V J184212.96-775807.0 : possible period changes
    \item[$\bullet$] V1356 Cyg : eccentric pair B
    \item[$\bullet$] Brh V154 : close to resonance 7:2 (only 0.1\% off)
    \item[$\bullet$] ASASSN-V J201545.10+373555.2 : eccentric pair A
    \item[$\bullet$] ZTF J205229.71+473345.9 : almost exactly resonance 5:3 (only 0.01\% off)
    \item[$\bullet$] ASASSN-V J222721.05+564425.3 : both pair A and B are eccentric
    \item[$\bullet$] ZTF J224132.79+582517.4 : eccentric pair B, longest period in our sample
    \item[$\bullet$] ASASSN-V J233336.79+615012.0 : eccentric pair A
\end{itemize}

One system, which definitely deserves our attention would be V0871 Cen (= HD 101205 = TIC 319936710).
This is the brightest system in our sample of stars, member of the cluster IC 2944, star very often
studied both photometrically, and spectroscopically (see e.g. \citealt{1992IBVS.3765....1M},
\citealt{2007OEJV...72....1O}, \citealt{2011MNRAS.416..817S}). Since now there was detected the
photometric period of about 2.09 days with its Algol-like eclipsing light curve, as well as the
spectroscopic period of 2.8 days. Moreover, there was found that the whole system is much more
complicated, containing also three other close visual companions besides the inner brightest A: B-C-D
(distant 0.36$^{\prime\prime}$, 1.7$^{\prime\prime}$, and 9.6$^{\prime\prime}$). However, it was not
known which of the inner ABC components contain the eclipsing binary and which the spectroscopic pair.
Up to now thanks to TESS. Using these super-precise data we were able to detect both 2.09 and 2.8 days
period signals in the photometric data, and moreover also to detect their possible mutual eclipse on
their orbit around a common barycenter. Due to the fact that the mutual movement of the A-B pair is
only very slow of the order of thousands of years \citep{2009AJ....138..664Z}, and hence the
probability of the mutual eclipse on such an orbit right now in the TESS epoch very improbable -- we
have to conclude that the architecture of the whole system is as follows. The most inner two pairs are
the 2.8 and 2.09 days binaries (we name it Aa-Ab), accompanied with a more distant component B with its
only poorly-constrained orbit by \cite{2009AJ....138..664Z}. And much more distant C and D components
are probably bound (due to their similar proper motion), but only very weakly. Hence the whole system
is likely a septuple one. The reason why its 2.8-d photometric variation was not noticed earlier is due
to its much lower amplitude compared to the dominant pair A.

\begin{figure}
  \centering
  \includegraphics[width=0.45\textwidth]{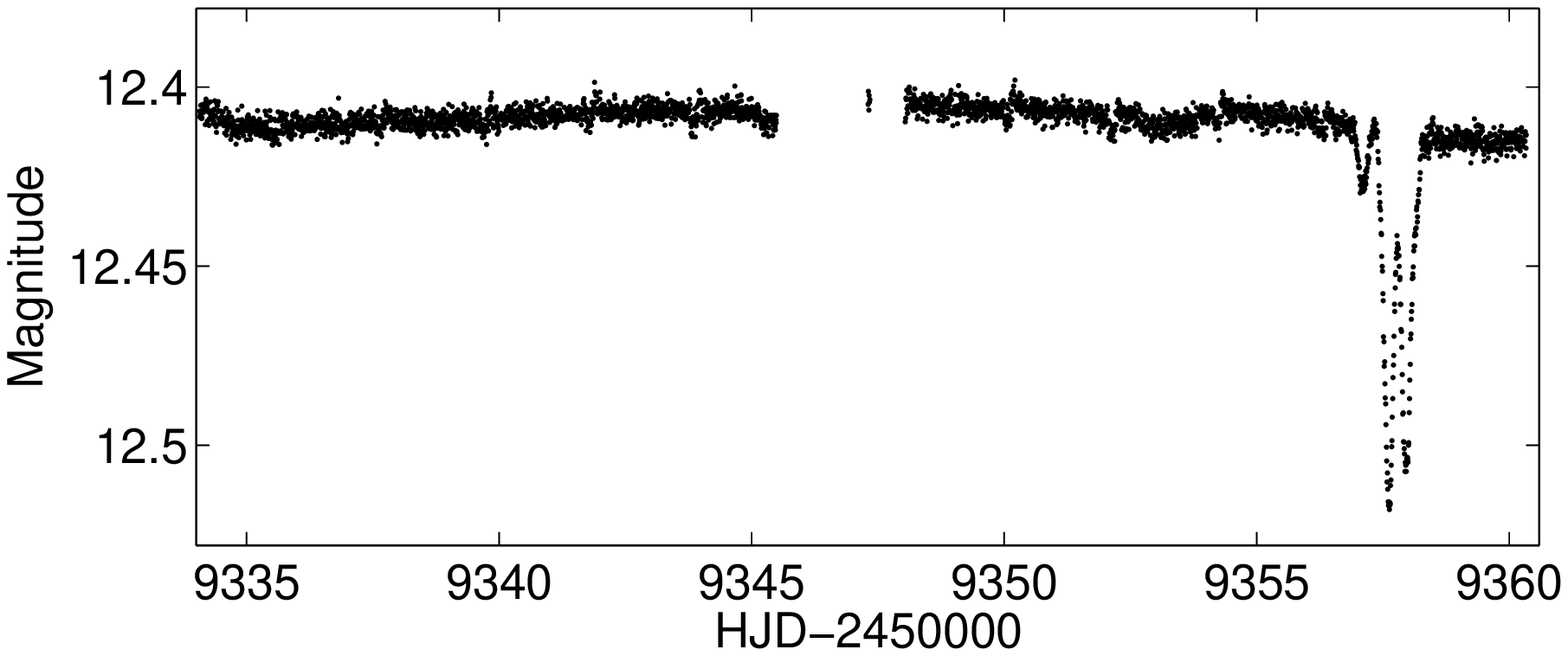}
  \caption{The system named ASASSN-V J124203.23-644513.2 showing besides the both inner eclipsing periods also additional eclipse with quite
  a complicated structure, indicating maybe a coplanar 2+2 system with its outer mutual eclipses. These data are 'pre-whitened' to clearly
  see this additional eclipse, i.e. both inner eclipsing curves were subtracted.}
  \label{ASAS124203}
 \end{figure}

And finally the last one system, which we found extremely interesting is the star named as ASASSN-V
J124203.23-644513.2. This system shows besides the two eclipsing periods of pairs A and B (periods
2.0725, and 1.4123 days) also an additional eclipse observed by TESS (see Fig.\ref{ASAS124203}). But
its shape is "double-peaked", like in triply eclipsing triples. And its depth is similar as for the
pair A, indicating that the pair A is being eclipsed at that time. It means three most probable
explanations emerge: \textcircled{\scriptsize 1} quadruple 2+2 doubly eclipsing system, which is also
perfectly coplanar that even this mutual orbit is eclipsing, or \textcircled{\scriptsize 2} doubly
eclipsing system with one additional component (architecture 2+2+1) causing this eclipse, or
\textcircled{\scriptsize 3} blend of stars not connected, hence two systems of a triple and a binary.





 \begin{table*}
   \caption{Doubly eclipsing candidates.}  \label{systemsInfoDEBs}
   \scalebox{0.52}{
   \begin{tabular}{c c c c c c | c c c c | c c c c}\\[-6mm]
 \hline \hline
              &              &                  &              &         &                  &             \multicolumn{4}{|c}{Pair A}                      & \multicolumn{4}{|c}{Pair B} \\
  RA [J2000.0]& DE [J2000.0] & VSX Target name  &  TESS number & EB type & Mag$_{max}$ $^\sharp$ & $JD_0-2450000$ & $P$[d] & $D_P$ [mag] & $D_S$ [mag] & $JD_0-2450000$ & $P$[d] & $D_P$ [mag] & $D_S$ [mag] \\[0.5mm]
  \hline
  00 02 46.39 & +69 52 24.78 &  WISE J000246.4+695224            & TIC 378584712 & EW + EA & 13.772 & 8988.052   & 0.2893554 & 0.1  & 0.1  & 8992.782   & 35.6614    & 0.3   & 0.25  \\
  00 04 32.60 & +63 26 05.03 &ASASSN-V J000432.60+632605.0       & TIC 283020555 & EA + EA & 12.332 & 8772.29400 & 1.0979473 & 0.4  & 0.5  & 8772.35990 &  0.9944210 & 0.1   & 0.08  \\
  00 47 27.28 & +64 49 04.87 &ASASSN-V J004727.28+644904.9 $^{*}$& TIC 284814380 & EA + EA & 12.230 & 8808.72977 & 4.0792904 & 0.22 & 0.13 & 8809.14576 &  4.9870291 & 0.07  & 0.04  \\
  00 59 40.89 & +61 11 15.32 & CzeV1435                          & TIC 256100687 & EA + EW & 15.684 & 8975.31886 & 1.7731144 & 0.20 & 0.16 & 8973.15154 & 0.3908587  & 0.04  & 0.04  \\
  01 33 10.90 & +61 35 07.04 &  WISE J013310.8+613507 $^{*}$     & TIC 389836747 & EA + EA & 10.800 & 8979.69150 & 2.5670368 & 0.23 & 0.18 & 8977.37517 &  2.7303226 & 0.032 & 0.016 \\
  02 03 06.68 & +62 43 15.42 &ASASSN-V J020306.68+624315.4       & TIC 400813988 & EA + EA & 13.022 & 8795.35118 & 3.28670   & 0.16 & 0.07 & 8799.01029 & 36.7607    & 0.3   & 0.14  \\
  02 21 05.82 & +59 55 16.9  &  ZTF J022105.82+595516.9          & TIC 357677349 & EB + EA & 14.509 & 8652.926   & 1.111972  & 0.14 & 0.10 & 8653.668   & 2.36685    & 0.2   & 0.03  \\
  02 42 21.82 & +62 54 03.64 &ASASSN-V J024221.82+625403.6       & TIC 50445777  & EA + EA & 13.695 & 8810.28712 & 5.7615    & 0.15 & 0.03 & 8791.580   & 11.5503    & 0.03  & 0.006 \\
  02 53 00.48 & +57 32 33.79 & V1361 Cas                         & TIC 251172666 & EA + EA & 13.399 & 8800.26400 & 2.2367206 & 0.2  & 0.2  & 8810.005   & 1.225857   & 0.014 & 0.013 \\
  03 01 19.37 & +60 34 20.24 & V1018 Cas $^{*}$                  & TIC 286470992 & EA + EA & 10.210 & 8821.08412 & 4.1277450 & 0.26 & 0.25 & 8820.14975 & 3.1052492  & 0.02  & 0.008 \\
  03 03 00.05 & +55 12 43.78 & ZTF J030300.04+551243.7           & TIC 251564581 & EA + EA & 14.628 & 8797.26242 & 7.7787461 & 0.07 & 0.07 & 8806.861   & 1.0125150  & 0.04  & 0.02  \\
  03 05 36.60 & +68 57 27.50 &ASASSN-V J030536.60+685727.5       & TIC 297696974 & EA + EA & 13.533 & 8823.39150 & 0.6022180 & 0.1  & 0.08 & 9007.5488  & 2.1660850  & 0.03  & 0.02  \\
  03 07 50.25 & +54 03 58.18 &ASASSN-V J030750.25+540358.2 $^{*}$& TIC 251757935 & EA + EA & 12.242 & 8798.787   & 1.5250    & 0.2  & 0.17 & 8805.35683 & 2.4076967  & 0.16  & 0.14  \\
  03 23 29.99 & +50 00 30.2  & ZTF J032329.99+500030.20          & TIC 252824398 & EA + EB & 14.130 & 8812.05    & 19.0003   & 0.42 & 0.10 & 8811.848   & 1.0102     & 0.07  & 0.04  \\
  04 11 07.45 & +56 22 32.99 & CzeV1254                          & TIC 260100848 & EA + EW & 13.304 & 8831.66206 & 0.7154189 & 0.13 & 0.12 & 8835.70322 & 0.3625926  & 0.1   & 0.08  \\
  04 58 56.56 & +67 46 55.63 & NSVS 2176810                      & TIC 468378722 & EA + EW & 13.517 & 8824.17492 & 3.6620862 & 1.4  & 0.13 & 8822.3560  & 0.4536163  & 0.18  & 0.17  \\
  05 13 31.78 & +35 39 11.02 & V0417 Aur $^{*}$                  & TIC 367448265 & EA + EW &  9.010 & 9507.10550 & 1.86553   & 0.26 & 0.22 & 9507.64033 & 0.4182123  & 0.04  & 0.03  \\
  05 21 21.25 & +34 38 05.13 & CzeV1493                          & TIC 2603042   & EA + EW & 12.686 & 9274.31558 & 1.9126763 & 0.24 & 0.08 & 9279.3310  & 0.2712981  & 0.03  & 0.03  \\
  05 34 08.96 & +35 17 50.10 & CzeV1759                          & TIC 115382018 & EA + EA & 12.992 & 8830.005   & 0.9456281 & 0.11 & 0.05 & 8829.86    & 3.4288     & 0.04  & 0.01  \\
  05 49 04.04 & +08 35 16.15 &ASASSN-V J054904.04+083516.2       & TIC 156501953 & EA + EA & 13.217 & 8483.67357 & 3.6274960 & 0.36 & 0.38 & 9209.80131 & 1.4772221  & 0.015 & 0.002 \\
  05 49 06.27 & +60 10 38.17 &  NSVS 2209083                     & TIC 70664819  & EA + EA & 10.469 & 8823.20330 & 2.6801419 & 0.18 & 0.18 & 8823.8330  & 0.5517664  & 0.006 & 0.002 \\
  05 54 02.99 & +01 40 21.81 & V1793 Ori                         & TIC 159089190 & EA + EA &  9.948 & 9208.691   & 3.55240   & 0.09 & 0.06 & 8473.452   & 3.3475421  & 0.06  & 0.06  \\
  06 14 13.85 & -07 07 54.16 &  WISE J061413.8-070754            & TIC 72555159  & EA + EA & 13.489 & 9206.78980 & 1.1867892 & 0.61 & 0.09 & 8487.30287 & 35.6518913 & 0.08  & 0.08  \\
  06 23 16.08 & +16 05 29.44 & ZTF J062316.07+160529.4           & TIC 438106145 & EA + EA & 14.127 & 8599.64900 & 0.6360014 & 0.09 & 0.05 & 9492.890   & 7.5366804  & 0.03  & 0.015 \\
  06 40 48.28 & -22 46 58.98 &ASASSN-V J064048.28-224659.0       & TIC 48614583  & EA + EA & 13.778 & 9218.7682  & 2.8505583 & 0.32 & 0.32 & 9220.60608 & 6.6412165  & 0.05  & 0.05  \\
  06 45 39.57 & +14 33 49.54 &ASASSN-V J064539.57+143349.6       & TIC 372429781 & EA + EA & 14.118 & 8484.54433 & 2.646248  & 0.35 & 0.07 & 9506.8150  & 4.8951198  & 0.06  & 0.03  \\
  06 50 36.70 & -03 41 30.30 & BEST-II lra2a\_00811              & TIC 281799398 & EA + EW & 13.421 & 8472.58734 & 1.8162916 & 0.07 & 0.04 & 8473.09032 & 0.3981756  & 0.04  & 0.04  \\
  06 56 16.66 & +15 48 32.54 & ZTF J065616.65+154832.5 $^{*}$    & TIC 386202029 & EB + EA & 13.522 & 9508.955   & 1.2513289 & 0.42 & 0.19 & 8485.94818 & 1.0600426  & 0.11  & 0.11  \\
  07 02 51.10 & +05 01 47.64 &ASASSN-V J070251.10+050147.6       & TIC 291497351 & EA + EA & 12.547 & 9217.98628 & 3.61330   & 0.43 & 0.38 & 9205.3290  & 1.4197927  & 0.014 & 0.014 \\
  07 06 05.88 & -03 00 26.57 & GDS\_J0706058-030026 $^{*}$       & TIC 177810207 & EA + EA & 13.546 & 8507.3370  & 1.4228503 & 0.19 & 0.15 & 9210.70294 & 1.737956   & 0.07  & 0.02  \\
  07 08 38.28 & -17 19 52.93 &ASASSN-V J070838.27-171952.9       & TIC 148942773 & EA + EA & 13.407 & 7042.794   & 4.3005671 & 0.13 & 0.11 & 9211.74592 & 3.7672138  & 0.04  & 0.03  \\
  07 11 31.63 & -15 33 41.26 &ASASSN-V J071131.63-153341.3       & TIC 306768262 & EA + EA & 12.685 & 9210.25430 & 2.611144  & 0.29 & 0.21 & 9208.837   & 1.7289645  & 0.015 & 0.007 \\
  07 23 04.90 & -11 00 43.52 &ASASSN-V J072304.90-110043.5       & TIC 187663930 & EA + EA & 12.325 & 9236.88043 & 2.553083  & 0.44 & 0.07 & 9233.43686 & 2.2011469  & 0.03  & 0.02  \\
  07 30 54.25 & -18 40 42.38 & ASAS J073054-1840.7               & TIC 456702203 & EA + EA & 12.272 & 9245.65345 & 2.068435  & 0.30 & 0.25 & 8508.99491 & 1.7285119  & 0.09  & 0.08  \\
  07 33 16.82 & -33 14 11.08 & BESTII F20a\_01769                & TIC 113385901 & EA + EW & 12.979 & 9238.77219 & 2.6519588 & 0.19 & 0.06 & 9235.29849 & 0.2897737  & 0.05  & 0.05  \\
  07 34 51.49 & -07 06 10.22 &ASASSN-V J073451.49-070610.2       & TIC 7215036   & EA + EA & 14.809 & 9246.00521 & 5.5879951 & 0.42 & 0.42 & 9234.9070  & 1.683838   & 0.03  & 0.06  \\
  07 35 15.34 & -38 45 50.98 &ASASSN-V J073515.34-384551.0       & TIC 174202524 & EA + EA & 14.324 & 9233.367   & 0.94730   & 0.25 & 0.17 & 8498.21238 & 1.1741810  & 0.15  & 0.14  \\
  07 38 36.07 & -25 41 10.68 &ASASSN-V J073836.07-254110.7       & TIC 110802531 & EA + EA & 14.310 & 9234.01097 & 6.3604401 & 0.65 & 0.12 & 9250.334   & 1.5345683  & 0.01  & 0.005 \\
  07 45 36.93 & -31 09 32.00 & V0674 Pup                         & TIC 149611278 & EW + EA & 10.869 & 8497.56967 & 0.6029032 & 0.31 & 0.23 & 8511.4347  & 6.5245459  & 0.29  & 0.29  \\
  07 48 44.44 & -37 49 59.95 &ASASSN-V J074844.43-374960.0       & TIC 130874723 & EA + EA & 12.537 & 9234.627   & 0.5055883 & 0.13 & 0.1  & 9236.85572 & 1.9735046  & 0.05  & 0.02  \\
  07 58 48.70 & -37 43 15.53 &  WISE J075848.7-374315            & TIC 132535542 & EW + EW & 14.809 & 9246.95869 & 0.395653  & 0.38 & 0.36 & 9236.48223 & 0.3093777  & 0.11  & 0.10  \\
  08 05 37.81 & +52 21 10.80 & LZ Lyn                            & TIC 252980672 & EB + EW & 10.163 & 8850.742   & 0.906131  & 0.95 & 0.38 & 9589.693   & 0.3956921  & 0.07  & 0.07  \\
  08 10 48.48 & +13 34 01.99 &ASASSN-V J081048.48+133402.0 $^{*}$& TIC 27543409  & EA + EA & 13.643 & 7283.1450  & 2.122740  & 0.07 & 0.02 & 9245.0180  & 4.0133499  & 0.08  & 0.02  \\
  08 15 49.78 & -36 07 50.09 &ASASSN-V J081549.78-360750.1       & TIC 182612964 & EA + EA & 13.128 & 9237.532   & 0.7775968 & 0.10 & 0.03 & 9261.415   & 5.0107097  & 0.07  & 0.07  \\
  08 26 16.93 & -67 54 27.25 &ASASSN-V J082616.93-675427.2       & TIC 307495225 & EA + EW & 12.352 & 9295.8365  & 1.4428125 & 0.1  & 0.04 & 9343.32143 & 0.3665626  & 0.04  & 0.04  \\
  08 46 01.42 & -37 02 23.35 &  WISE J084601.4-370223            & TIC 181567057 & EW + EW & 12.923 & 9259.51295 & 0.3151640 & 0.09 & 0.09 & 9260.10434 & 0.3838160  & 0.02  & 0.02  \\
  08 49 51.41 & -34 45 33.84 & CPD-34 3002                       & TIC 186379217 & EA + EA & 10.458 & 9260.2598  & 2.3185660 & 0.16 & 0.08 & 9259.6533  & 2.7375104  & 0.01  & 0.005 \\
  08 50 25.05 & -09 30 57.49 & NSVS 15567664                     & TIC 13539679  & EB + EW & 12.730 & 8513.918   & 0.5898351 & 0.31 & 0.18 & 9238.0685  & 0.3810852  & 0.08  & 0.08  \\
  08 56 31.48 & -40 14 12.59 &ASASSN-V J085631.48-401412.6 $^{*}$& TIC 190895730 & EA + EA & 12.482 & 8551.011   & 1.31790   & 0.08 & 0.06 & 8551.62591 & 0.9182636  & 0.13  & 0.11  \\
  09 19 51.17 & -59 33 06.90 &ASASSN-V J091951.17-593306.9       & TIC 386900081 & EA + EA & 13.825 & 9314.28181 & 2.6686687 & 0.3  & 0.2  & 9315.64518 & 14.33574   & 0.25  & 0.1   \\
  09 20 31.34 & -54 24 38.09 &ASASSN-V J092031.34-542438.1 $^{*}$& TIC 387096013 & EA + EA & 12.185 & 9273.55373 & 4.268089  & 0.21 & 0.17 & 9263.87932 & 6.1442495  & 0.08  & 0.05  \\
  09 54 53.84 & -72 04 36.92 &ASASSN-V J095454.54-720436.1       & TIC 453335440 & EA + EW & 13.427 & 9342.184   & 0.7465108 & 0.25 & 0.25 & 9323.5446  & 0.3634727  & 0.12  & 0.12  \\
  10 08 20.04 & -73 15 54.00 & WISE J100820.0-731554             & TIC 453459782 & EA + EA & 14.922 & 9312.38265 & 0.3683176 & 0.19 & 0.19 & 9338.69558 & 7.9453491  & 0.31  & 0.25  \\
  10 29 11.57 & -52 24 13.57 &ASASSN-V J102911.57-522413.6       & TIC 447369043 & EA + EA & 13.358 & 9312.51785 & 0.5727205 & 0.11 & 0.05 & 9314.99332 & 3.7902466  & 0.10  & 0.05  \\
  10 34 48.55 & -60 13 03.90 & ASAS J103449-6013.1               & TIC 851090533 & EA + EA & 9.718  & 9309.7709  & 7.593710  & 0.27 & 0.24 & 8616.1536  & 3.1820930  & 0.02  & 0.005 \\
  10 42 06.48 & -42 52 40.69 & V0432 Vel  $^{*}$                 & TIC 146810480 & EA + EA & 9.445  & 9288.81961 & 1.468570  & 0.15 & 0.1  & 9288.528   & 0.5449716  & 0.03  & 0.02  \\
  10 52 24.11 & -55 47 46.68 &ASASSN-V J105224.11-554746.7       & TIC 459601989 & EA + EA & 12.250 & 9296.310   & 1.4961114 & 0.14 & 0.05 & 9287.28115 & 53.1092318 & 0.1   & 0.08  \\
  10 58 24.33 & -61 13 47.57 &ASASSN-V J105824.33-611347.6       & TIC 465899856 & EA + EA & 12.673 & 8586.46522 &12.9954333 & 0.14 & 0.1  & 9317.2230  & 2.3307278  & 0.03  & 0.015 \\
  11 26 01.23 & -66 11 28.75 &ASASSN-V J112601.23-661128.8       & TIC 295767114 & EA + EA & 14.941 & 9343.04102 & 2.8053980 & 0.34 & 0.23 & 9341.2973  & 1.1937377  & 0.05  & 0.01  \\
  11 34 26.20 & -63 20 02.00 & ASAS J113426-6320.0               & TIC 319121329 & EA + EA & 10.867 & 9340.05011 & 1.7172011 & 0.18 & 0.11 & 9343.35465 & 5.0107072  & 0.12  & 0.04  \\
  11 38 20.36 & -63 22 21.89 & V0871 Cen                         & TIC 319936710 & EA + EW &  6.500 & 8992.9445  & 2.090704  & 0.08 & 0.08 & 8994.30    & 2.8170     & 0.02  & 0.015 \\
  11 39 17.31 & -62 10 29.21 & V0384 Cen                         & TIC 320239566 & EA + EA & 11.59  & 3144.7030  & 12.635401 & 0.45 & 0.36 & 8606.15859 & 3.6402566  & 0.05  & 0.03  \\
  11 42 06.66 & -62 16 02.21 &  [DPS2013] 60 $^{*}$              & TIC 321471064 & EA + EA & 12.487 & 9340.525   & 0.6842184 & 0.15 & 0.15 & 9341.53809 & 1.5572506  & 0.15  & 0.04  \\
  11 52 25.73 & -66 57 54.22 & WISE J115225.7-665754             & TIC 410922760 & EA + EA & 12.218 & 9344.56349 & 2.0461214 & 0.21 & 0.15 & 9349.5327  & 3.3214941  & 0.04  & 0.02  \\
  12 15 29.02 & -62 39 28.66 & KELT KS38C016096                  & TIC 383338471 & EA + EA & 10.782 & 9318.4924  & 2.7914930 & 0.07 & 0.01 & 9309.8392  & 31.57253   & 0.05  & 0.05  \\
  12 42 03.23 & -64 45 13.14 &ASASSN-V J124203.23-644513.2       & TIC 433545934 & EA + EA & 13.391 & 9352.210   & 2.0725413 & 0.31 & 0.29 & 9341.77135 & 1.4122993  & 0.1   & 0.06  \\
  12 54 27.31 & -65 34 37.67 &ASASSN-V J125427.31-653437.7       & TIC 436598714 & EA + EA & 13.513 & 8194.796   & 1.8401023 & 0.16 & 0.1  & 9340.76110 & 1.8893375  & 0.04  & 0.02  \\
  13 47 43.51 & -51 24 16.74 &ASASSN-V J134743.51-512416.7       & TIC 241564994 & EB + EB & 12.868 & 7767.793   & 0.5774803 & 0.23 & 0.13 & 9350.87648 & 0.4939062  & 0.05  & 0.03  \\
  14 07 54.66 & +35 44 56.08 &ASASSN-V J140754.66+354456.0       & TIC 138186185 & EA + EA & 13.341 & 8932.992   & 1.0129093 & 0.27 & 0.25 & 8934.379   & 8.2610749  & 0.08  & 0.03  \\
  14 41 24.41 & -71 02 53.20 & HD 128523 $^{*}$                  & TIC 257776944 & EA + EA &  9.861 & 9367.32122 & 2.4510600 & 0.19 & 0.16 & 9369.79461 & 3.3045126  & 0.05  & 0.02  \\
  14 35 36.01 & -72 14 59.46 &ASASSN-V J143536.01-721459.4       & TIC 401924368 & EA + EA & 12.710 & 8642.16487 & 7.3536094 & 0.42 & 0.37 & 9379.33724 & 6.368810   & 0.03  & 0.03  \\
  15 40 12.48 & -72 34 45.98 &ASASSN-V J154012.48-723446.0       & TIC 425860943 & EA + EA & 12.583 & 9368.94446 & 1.054818  & 0.21 & 0.04 & 9371.59362 & 0.4812192  & 0.02  & 0.02  \\
  15 40 43.29 & -49 48 08.35 &ASASSN-V J154043.29-494808.4       & TIC 190149246 & EA + EA & 12.708 & 8216.812   & 2.6953877 & 0.22 & 0.19 & 8647.64254 & 0.5670568  & 0.09  & 0.08  \\
  15 49 58.67 & -49 48 17.60 &ASASSN-V J154958.67-494817.6       & TIC 276728788 & EA + EW & 13.497 & 9372.9290  & 6.0683920 & 1.61 & 0.18 & 9379.3882  & 0.3874346  & 0.06  & 0.06  \\
  15 51 57.55 & -43 05 47.72 &ASASSN-V J155157.55-430547.7       & TIC 254963234 & EA + EA & 11.217 & 8802.00336 & 2.92193   & 0.11 & 0.11 & 8802.363   & 8.931      & 0.04  & 0.04  \\
  16 04 11.50 & +43 01 49.08 &ASASSN-V J160411.50+430149.1 $^{*}$& TIC 219469945 & EA + EA & 12.618 & 8992.31791 & 2.717870  & 0.15 & 0.05 & 8972.66544 & 14.9637727 & 0.08  & 0.02  \\
  16 16 01.84 & -42 45 33.05 &ASASSN-V J161601.84-424533.0       & TIC 14700313  & EA + EW & 14.609 & 9373.247   & 3.2005508 & 0.25 & 0.23 & 9372.4670  & 0.3638294  & 0.09  & 0.09  \\
  16 21 33.00 & -67 59 19.07 &ASASSN-V J162133.01-675919.1       & TIC 362322922 & EA + EW & 13.859 & 9370.78755 & 2.3832433 & 1.3  & 0.3  & 9369.6571  & 0.4278364  & 0.05  & 0.05  \\
  16 21 56.98 & -49 09 24.48 &  V0398 Nor                        & TIC 216088118 & EA + EA &  9.576 & 9371.93919 & 1.588950  & 0.3  & 0.2  & 9370.12706 & 1.5410346  & 0.05  & 0.02  \\
  16 27 38.80 & -43 29 23.68 &ASASSN-V J162738.80-432923.7       & TIC 225009475 & EW + EW & 12.646 & 9370.67464 & 0.5375934 & 0.24 & 0.19 & 9369.60617 & 0.4115693  & 0.07  & 0.07  \\
  16 56 56.95 & +29 19 06.49 &  V1037 Her                        & TIC 286280289 & EA + EA & 11.910 & 8992.62367 & 0.7875805 & 0.3  & 0.19 & 8994.20931 & 5.803712   & 0.1   & 0.01  \\
  17 10 20.36 & -79 40 22.98 &ASASSN-V J171020.36-794023.0       & TIC 384707815 & EB + EA & 13.395 & 9371.561   & 0.5197554 & 0.19 & 0.11 & 8662.39568 & 1.4566890  & 0.19  & 0.06  \\
  17 25 57.86 & -70 25 16.03 &ASASSN-V J172557.86-702516.0       & TIC 293328230 & EA + EW & 12.801 & 9370.59528 & 4.3982343 & 0.3  & 0.26 & 9370.90124 & 0.324594   & 0.05  & 0.04  \\
  17 33 44.14 & -36 30 37.80 &ASASSN-V J173344.14-363037.8       & TIC 465096122 & EA + EA & 13.047 & 7893.810   & 5.4788615 & 0.29 & 0.07 & 9371.93765 & 2.6152141  & 0.08  & 0.07  \\
  17 50 00.12 & -41 52 47.89 &ASASSN-V J175000.12-415247.9       & TIC 21369338  & EA + EA & 12.41  & 9370.69407 & 3.2902183 & 0.18 & 0.16 & 9369.7630  & 3.7108779  & 0.03  & 0.01  \\
  17 59 44.96 & -50 21 22.82 &  WISE J175944.9-502122            & TIC 381608215 & EA + EA & 15.93  & 8662.1669  & 0.374864  & 0.1  & 0.1  & 9380.2610  & 1.0160054  & 0.05  & 0.03  \\
  18 24 00.66 & -35 42 10.94 &ASASSN-V J182400.66-354210.9       & TIC 66416271  & EA + EW & 14.80  & 8673.12504 & 3.3017214 & 0.4  & 0.25 & 8664.32946 & 0.37220    & 0.1   & 0.1   \\
  18 29 34.18 & -39 50 10.90 & WISE J182934.1-395010             & TIC 312653568 & EW + EB & 12.863 & 8675.96475 & 0.5913301 & 0.36 & 0.36 & 8676.03176 & 0.5383272  & 0.04  & 0.03  \\
  18 41 17.47 & -41 40 37.38 &ASASSN-V J184117.47-414037.4 $^{*}$& TIC 316464966 & EA + EA & 13.962 & 8011.565   & 2.124815  & 0.15 & 0.1  & 8666.4356  & 0.4845352  & 0.05  & 0.05  \\
  18 42 12.96 & -77 58 07.03 & ASASSN-V J184212.96-775807.0      & TIC 351918052 & EW + EW & 13.808 & 9369.31215 & 0.5412306 & 0.17 & 0.17 & 9371.117   & 0.3405923  & 0.07  & 0.06  \\
  18 47 00.16 & -49 22 55.49 & ASASSN-V J184700.16-492255.5      & TIC 158159279 & EA + EA & 12.379 & 8666.83878 & 4.1765513 & 0.15 & 0.15 & 8667.9695  & 3.2577395  & 0.03  & 0.028 \\
  19 10 22.62 & +16 43 03.18 & ASASSN-V J191022.62+164303.2      & TIC 396181925 & EA + EA & 12.238 & 9407.13905 & 1.20379   & 0.31 & 0.27 & 9401.99234 & 1.3815730  & 0.07  & 0.025 \\
  19 19 51.11 & +34 58 15.24 &  KELT KC11C034703                 & TIC 392799977 & EA + EA & 11.910 & 8702.02833 & 2.1919011 & 0.03 & 0.005& 8700.58204 & 9.3813864  & 0.026 & 0.026 \\
  19 22 23.44 & +36 17 39.0  &  CzeV2708                         & TIC 122513089 & EA + EW & 13.217 & 9337.50142 & 1.24205   & 0.5  & 0.5  & 8688.2536  & 0.2726330  & 0.03  & 0.03  \\
  19 43 02.90 & +29 48 14.08 & ASASSN-V J194302.90+294814.1      & TIC 285286180 & EA + EA & 12.374 & 9443.8350  & 1.6314193 & 0.13 & 0.08 & 9426.8663  & 3.3814447  & 0.03  & 0.02  \\
  19 48 56.48 & +36 03 09.25 & ZTF J194856.47+360309.2           & TIC 169282552 & EA + EA & 15.160 & 8687.63587 & 0.6413882 & 0.29 & 0.11 & 8700.19618 & 0.8664298  & 0.19  & 0.06  \\
  19 56 39.41 & +29 59 29.18 &  V1356 Cyg                        & TIC 87926550  & EW + EA & 10.305 & 8688.20332 & 1.9566675 & 0.28 & 0.25 & 9430.99117 & 7.5682323  & 0.15  & 0.14  \\
  19 57 56.55 & +27 19 12.86 &  ZTF J195756.55+271912.8          & TIC 282890225 & EA + EA & 14.297 & 9430.000   & 2.8367388 & 0.16 & 0.07 & 9437.21991 & 3.5676917  & 0.03  & 0.03  \\
  20 00 52.75 & +30 52 05.66 &  ZTF J200052.75+305205.6          & TIC 103990287 & EA + EA & 13.921 & 8280.831   & 19.9998494& 0.35 & 0.21 & 9428.28247 & 3.4183586  & 0.02  & 0.02  \\
  20 06 42.53 & +24 59 20.72 &  Brh V154                         & TIC 244645340 & EW + EA & 13.013 & 9430.42216 & 0.5549471 & 0.25 & 0.25 & 9439.46272 & 1.9445266  & 0.10  & 0.07  \\
  20 15 45.10 & +37 35 55.18 & ASASSN-V J201545.10+373555.2      & TIC 11479549  & EA + EA & 14.165 & 8729.83    & 2.75347   & 0.28 & 0.22 & 8725.19201 & 2.8234774  & 0.04  & 0.03  \\
  20 17 42.81 & +36 28 21.25 &  ZTF J201742.81+362821.2          & TIC 1967314693& EA + EW & 15.566 & 9428.990   & 1.1787963 & 0.35 & 0.12 & 9426.7700  & 0.6853393  & 0.05  & 0.04  \\
  20 22 44.14 & +32 52 18.70 &  WISE J202244.1+325218            & TIC 136532215 & EA + EA & 16.165 & 9428.04542 & 3.4272801 & 0.47 & 0.33 & 8731.29995 & 21.471980  & 0.11  & 0.11  \\
  20 38 52.00 & +50 28 00.48 &  WISE J203851.9+502800 $^{*}$     & TIC 322727163 & EA + EA & 11.001 & 9427.35262 & 1.6402475 & 0.24 & 0.18 & 9429.98542 & 1.1562320  & 0.036 & 0.01  \\
  20 52 24.55 & +38 10 18.30 &  NSVS 5871089                     & TIC 194899489 & EB + EW & 11.100 & 9440.86459 & 0.7480267 & 0.35 & 0.17 & 9426.563   & 0.5378686  & 0.10  & 0.09  \\
  20 52 29.72 & +47 33 45.94 &  ZTF J205229.71+473345.9          & TIC 356664286 & EA + EW & 13.838 & 8728.4270  & 1.922342  & 0.52 & 0.14 & 8715.64347 & 1.153058   & 0.11  & 0.07  \\
  21 02 30.88 & +61 08 16.66 &  WISE J210230.8+610816            & TIC 305635022 & EA + EB & 14.729 & 8965.04512 & 1.8432113 & 0.2  & 0.17 & 8964.1850  & 0.571602   & 0.055 & 0.035 \\
  21 04 45.64 & +49 50 05.89 &  ZTF J210445.64+495005.8          & TIC 290232009 & EA + EA & 14.622 & 8271.948   & 1.2773286 & 0.26 & 0.16 & 8744.24421 & 3.4406492  & 0.12  & 0.07  \\
  21 21 28.81 & +51 48 55.62 &  ZTF J212128.80+514855.6          & TIC 63701935  & EW + EA & 13.90  & 8719.3930  & 2.195314  & 0.57 & 0.57 & 8746.23559 & 1.6280104  & 0.08  & 0.08  \\
  21 38 04.77 & +28 10 07.79 & ASASSN-V J213804.77+281007.8      & TIC 298712807 & EA + EA & 12.540 & 8728.15453 & 0.7472450 & 0.13 & 0.12 & 8728.79408 & 2.083481   & 0.055 & 0.04  \\
  22 04 17.14 & +55 26 07.30 &  ZTF J220417.14+552607.2          & TIC 470026602 & EA + EA & 13.546 & 8741.69694 & 1.772187  & 0.035& 0.035& 8754.79344 & 1.4225802  & 0.03  & 0.026 \\
  22 05 18.78 & +59 26 42.11 &  ZTF J220518.78+592642.1 $^{*}$   & TIC 327885074 & EA + EA & 13.551 & 8770.79968 & 2.7956986 & 0.23 & 0.2  & 8961.03205 & 3.3460783  & 0.04  & 0.01  \\
  22 27 21.05 & +56 44 25.30 &ASASSN-V J222721.05+564425.3 $^{*}$& TIC 414026507 & EA + EA & 11.151 & 8781.20497 & 6.4564804 & 0.24 & 0.17 & 8759.41453 & 4.2290212  & 0.13  & 0.11  \\
  22 41 32.79 & +58 25 17.40 &  ZTF J224132.79+582517.4          & TIC 388784863 & EA + EA & 14.233 & 8772.67696 & 1.1628068 & 0.13 & 0.11 & 8739.295   & 119.7988   & 0.35  & 0.32  \\
  23 02 00.88 & +40 58 40.19 &  WISE J230200.8+405840            & TIC 373874687 & EA + EW & 12.737 & 8755.95419 & 0.996781  & 0.51 & 0.19 & 8757.59412 & 0.3065449  & 0.02  & 0.01  \\
  23 33 36.79 & +61 50 12.05 & ASASSN-V J233336.79+615012.0      & TIC 270920077 & EA + EA & 11.146 & 8975.60741 & 3.3427872 & 0.14 & 0.12 & 8799.73119 & 1.2667572  & 0.04  & 0.03  \\
  \hline
 \end{tabular}} \\
  {\small Notes: $^\sharp$ - Out-of-eclipse magnitude, $V_{mag}$ from UCAC4 catalogue \citep{2013AJ....145...44Z}, or Guide Star Catalog II
  \citep{2008AJ....136..735L}, \\ $^{*}$ - Independently discovered during preparation of the current manuscript, and recently published in \cite{2022ApJS..259...66K}. The columns
  $D_P$, and $D_S$ denote the approximate depths of primary and secondary eclipses of both pairs based on the TESS data.}
 \end{table*}

 \begin{table*}
   \caption{Triply eclipsing triple candidates.}  \label{systemsInfoTriples}
   \scalebox{0.6}{
   \begin{tabular}{c c c c c | c c c c | c | l }\\[-6mm]
 \hline \hline
              &              &                                     &               &                   &           \multicolumn{4}{|c|}{Pair A}               & \multicolumn{2}{|c}{Outer orbit}    \\
  RA [J2000.0]& DE [J2000.0] & VSX Target name                     & TESS number   &  Mag$_{max}$ $^\sharp$ & $JD_0-2450000$ & $P$[d] & $D_P$ [mag] & $D_S$ [mag] & $JD_0-2450000$ & Comment \\[0.5mm]
  \hline
 04 31 15.65 & +57 45 01.12 & ASASSN-V J043115.65+574501.1         & TIC 356324779 & 13.229 & 8833.09262 & 3.476856  & 0.28  & 0.28  & 8837.5  & 86.693 d outer period, eccentric \\
 06 26 37.57 & -03 23 50.64 & ASASSN-V J062637.57-032350.6 $^{**}$ & TIC 42565581  & 13.714 & 9210.994   & 1.823075  & 0.15  & 0.15  & 9219.67 & 123.48 d outer period \\
 07 02 36.10 & -15 15 46.40 & WISE J070236.0-151546                & TIC 147911867 & 15.351 & 9219.77689 & 0.350096  & 0.29  & 0.29  & 9212.30 & \\
 07 34 10.87 & +04 19 23.81 & ASASSN-V J073410.87+041923.8         & TIC 321078998 & 14.211 & 9250.73878 & 3.707523  & 0.15  & 0.12  & 9334.2  & 202.09 d outer period \\
 09 43 29.56 & -43 06 48.89 & ASASSN-V J094329.56-430648.9         & TIC 33825824  & 13.900 & 9287.65065 & 9.230     & 0.3   & 0.25  & 8546.15 & \\
 10 12 37.44 & -59 43 44.80 & ASASSN-V J101237.44-594344.8         & TIC 463210472 & 12.719 & 9290.50694 & 5.6666999 & 0.11v & 0.09v & 8572.32 & 132.51 d outer period, eccentric \\
 10 39 08.91 & -53 55 45.59 & ASASSN-V J103908.91-535545.6         & TIC 303409465 & 13.393 & 7958.469   & 1.4261808 & 0.25  & 0.025 & 9289.42 & \\
 11 10 03.23 & -61 01 43.61 & OGLE CAR\_SC3\_98699                 & TIC 467049289 & 14.238 & 9315.45321 & 3.292551  & 0.24  & 0.24  & 9308.78 & 39.37 d outer period \\
 11 43 31.07 & -59 48 11.95 & ASASSN-V J114331.07-594811.9         & TIC 267413009 & 12.880 & 8607.33178 & 2.57630   & 0.13  & 0.05  & 9349.45 & \\
 11 57 59.86 & -63 55 55.45 & ASASSN-V J115759.86-635555.4         & TIC 306801572 & 11.421 & 9340.33937 & 5.5989301 & 0.17  & 0.17  & 8607.51 & \\
 12 03 45.72 & -62 38 53.59 & WX Cru                               & TIC 379344774 & 13.780 & 9344.08598 & 0.8375748 & 0.23  & 0.12  & 8604.05 &  60.27 d outer period, very deep \\
 12 22 36.55 & -12 53 57.48 & ASASSN-V J122236.55-125357.5         & TIC 349124978 & 12.805 & 8591.06132 & 9.872767  & 0.33  & 0.28  & 8573.08 &  \\
 12 38 53.66 & +31 56 59.78 & KELT KC08C11210                      & TIC 376606423 & 11.169 & 8907.6040  & 0.8547895 & 0.01  & 0.005 & 8908.09 &  \\
 14 51 09.98 & -63 35 08.77 & ASASSN-V J145109.98-633508.8         & TIC 294803663 & 12.457 & 9371.13272 & 2.2460229 & 0.28  & 0.21  & 8629.19 & 153.1 d outer period \\
 17 04 25.49 & +46 35 33.58 & CSS\_J170425.5+463533                & TIC 198581208 & 14.513 & 8986.3775  & 2.876069  & 0.19  & 0.02  & 9013.0  & 37.78 d outer period \\
 17 13 43.81 & -33 04 06.89 & V0726 Sco                            & TIC 47151245  & 10.349 & 9377.36748 & 1.20179   & 0.07  & 0.07  & 9368.49 & 285 d period, eccentric \\
 18 14 41.19 & -75 55 30.22 & ASASSN-V J181441.19-755530.2         & TIC 343763382 & 11.714 & 9380.99839 & 5.8662    & 0.39  & 0.20  & 8326.00 & 148.9 d outer period \\
 18 54 44.67 & +68 46 50.66 & WISE J185444.6+684650                & TIC 229785001 & 11.856 & 8935.38023 & 0.9295082 & 0.14  & 0.07  & 9402.11 & 165.28 d outer period ? \\
 19 50 09.98 & +41 57 03.31 & KIC 06543674                         & TIC 273231378 & 13.28  & 8700.09162 & 2.391047  & 0.6   & 0.57  & 9426.75 & known coplanar triple, see \citep{2015ApJ...806L..37M} \\
 20 17 00.33 & +39 08 19.61 & V0699 Cyg                            & TIC 11918748  & 11.598 & 9436.56753 & 3.10304   & 0.08  & 0.08  & 9423.86 & \\
 20 27 24.12 & +57 22 12.36 & KELT KC24C015704                     & TIC 467220877 & 11.479 & 8746.01890 & 5.6843594 & 0.03  & 0.028 & 8720.38 & \\
 21 52 19.89 & +55 21 56.92 & ZTF J215219.88+552156.9              & TIC 331266974 & 14.342 & 8782.29990 & 1.1989981 & 0.28  & 0.05  & 8747.48 & \\
 22 19 19.64 & +85 04 13.37 & ASASSN-V J221919.64+850413.4         & TIC 461500036 & 14.47  & 9001.44139 & 2.4711883 & 0.66  & 0.63  & 9005.60 & 54.27 d outer orbit?, eccentric \\
 22 37 38.49 & +61 22 09.59 & WISE J223738.5+612209                & TIC 455859832 & 14.59  & 8746.22562 & 1.1016521 & 0.2   & 0.16  & 8974.95 & 61.1 d outer orbit ? \\
 23 25 56.49 & +60 24 40.64 & ZTF J232556.49+602440.6              & TIC 270134953 & 14.56  & 8962.83469 & 1.3227165 & 0.19  & 0.06  & 8972.49 & \\
  \hline
 \end{tabular}}\\[1mm]
  {\scriptsize Notes: $^\sharp$ - Out-of-eclipse magnitude, $^{**}$ - Independently discovered during preparation of the current manuscript, and recently published in \cite{Arxiv2022}}
 \end{table*}

  \begin{figure}
  \centering
  \includegraphics[width=0.43\textwidth]{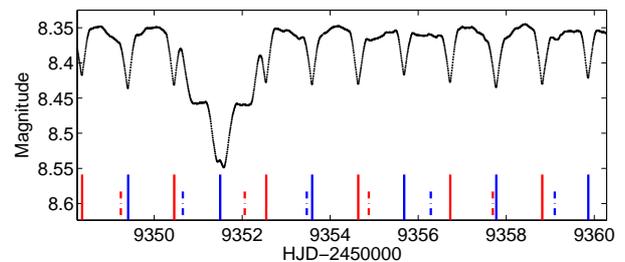}
  \caption{The system V0871 Cen with its complete light curve showing besides two prominent periods P$_{Aa}$ and P$_{Ab}$ (eclipses of both pairs denoted by short
  abscissae of blue colour for primary, and red for secondary eclipses, longer solid ones for pair Aa, while shorter dash-dotted for pair Ab)
  also an additional eclipse, which we believe is the outer eclipse of their mutual orbit Aa-Ab around a common barycenter. }
  \label{V871Cen}
 \end{figure}

\section{Conclusions}

We have carried out an analysis of all EA-type binaries from the VSX database (with $<$15 mag) trying
to identify there the additional eclipses in the TESS data. Our compilation of altogether 141 systems
is far the most extensive among other similar studies. Our presented database should serve the
observers and keen astronomers for detailed follow-up monitoring of these interesting targets. Due to
this reason we gave the ephemerides for both eclipsing pairs, but also their eclipse depths. These are
the crucial parameters for the prospective future observations.

Such a monitoring is potentially very important due to a chance of detecting the ETV for both pairs and
hence prove their quadruple nature. According to our experience \citep{2019A&A...630A.128Z} we should
believe that most of these candidate stars will be confirmed as quadruples due to their intensive
ground-based observations in the following years. Such a long-lasting monitoring would be our main task
for the future seasons. We already started collecting the data with our group of keen astronomers using
their quite modest technique, but still adequate for such a task. We found a combination of two
softwares SIPS (for reduction of the CCD frames) + SILICUPS (for plotting and subtracting the
individual light curve shapes) as very suitable for reducing and analysis of even such complicated
systems like the doubly eclipsing systems are.

\medskip

\underline{Note added in proof: } Just after the submission of the present manuscript there was
published an independent paper dealing with the same topic, \cite{2022ApJS..259...66K}. The authors
presented a group of 97 doubly eclipsing systems found in the TESS database, but identified with
different method (scanning all stars instead of only known eclipsing binaries like we did). Due to this
reason there was only quite small overlap of both our groups of stars, in total 18 systems marked with
asterisk in Table \ref{systemsInfoDEBs}.

\begin{acknowledgements}
We would like to thank Dr. Dariusz Graczyk as a referee for his valuable suggestions and remarks
improving the overall quality of the manuscript.
 We do thank the NSVS, ZTF, ASAS-SN, SWASP, and TESS teams for making all of the observations easily
public available. The research of P.Z. was supported by the project Progress Q47 {\sc Physics} of the
Charles University in Prague. This work is supported by MEYS (Czech Republic) under the project MEYS
LTT17006. This work has made use of data from the European Space Agency (ESA) mission {\it Gaia}
(\url{https://www.cosmos.esa.int/gaia}), processed by the {\it Gaia} Data Processing and Analysis
Consortium (DPAC, \url{https://www.cosmos.esa.int/web/gaia/dpac/consortium}). Funding for the DPAC has
been provided by national institutions, in particular the institutions participating in the {\it Gaia}
Multilateral Agreement. This research made use of Lightkurve, a Python package for TESS data analysis
\citep{2018ascl.soft12013L}. This research has made use of the SIMBAD and VIZIER databases, operated at
CDS, Strasbourg, France and of NASA Astrophysics Data System Bibliographic Services.
\end{acknowledgements}

 \begin{appendix}
 \section{Appendix A: Additional material}

 \begin{figure*}
  \centering
  \includegraphics[width=0.90\textwidth]{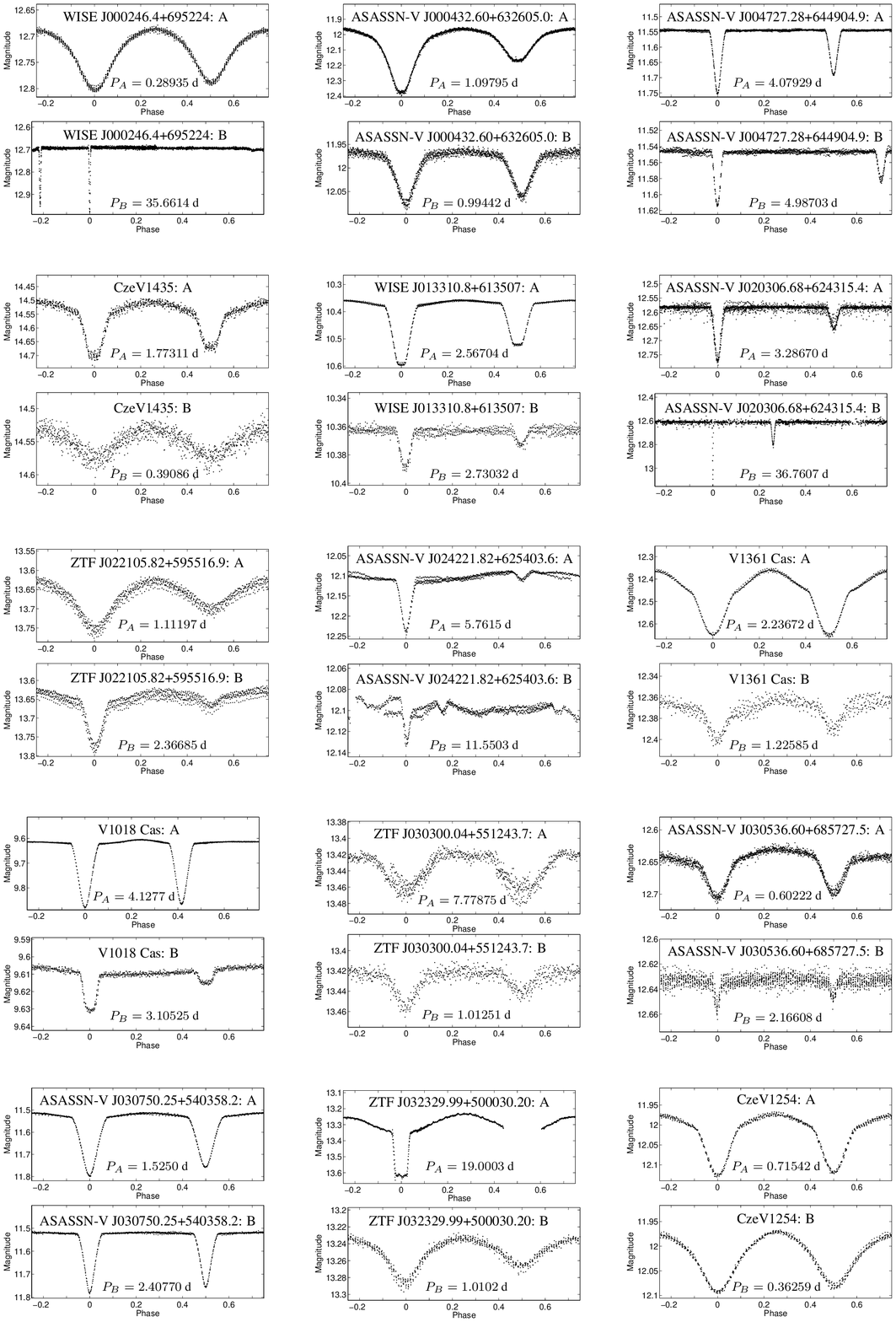}
  \caption{The light curves of both pairs as disentangled from the original TESS photometry. Plotted with ascending order of right ascension.}
  \label{LC01}
 \end{figure*}

  \begin{figure*}
  \centering
  \includegraphics[width=0.90\textwidth]{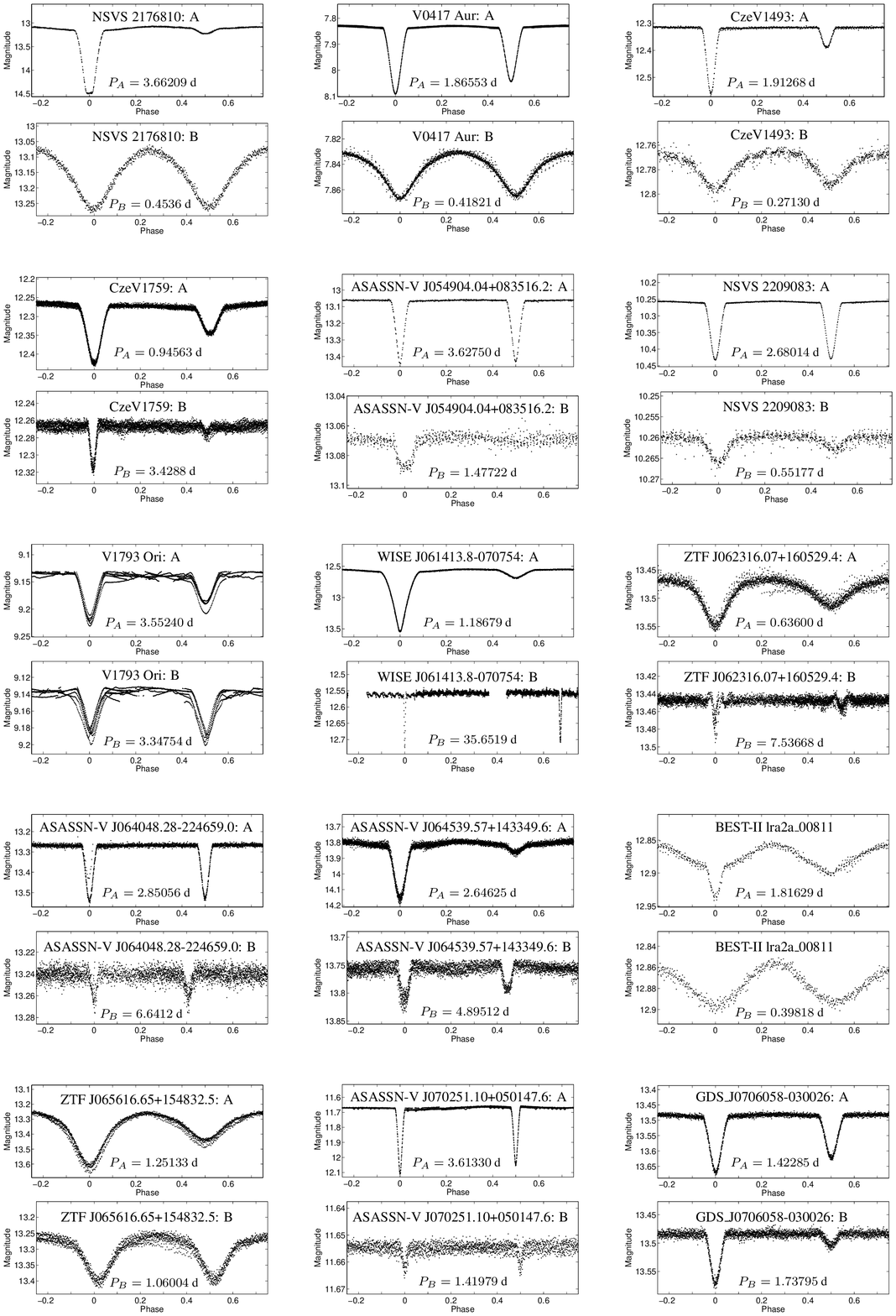}
  \caption{The light curves of both pairs as disentangled from the original TESS photometry, continuation.}
  \label{LC02}
 \end{figure*}

  \begin{figure*}
  \centering
  \includegraphics[width=0.90\textwidth]{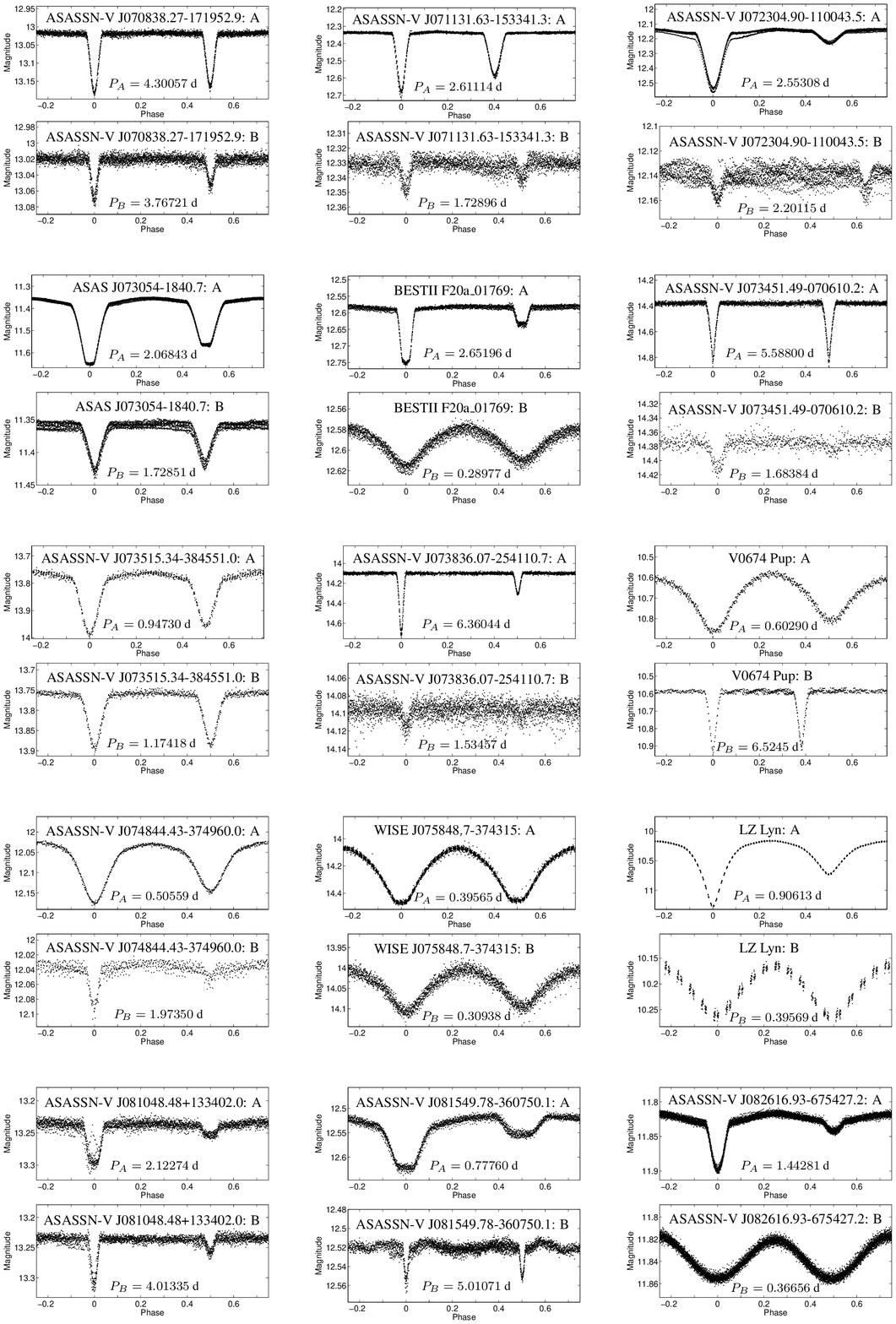}
  \caption{The light curves of both pairs as disentangled from the original TESS photometry, continuation.}
  \label{LC03}
 \end{figure*}

 \begin{figure*}
  \centering
  \includegraphics[width=0.90\textwidth]{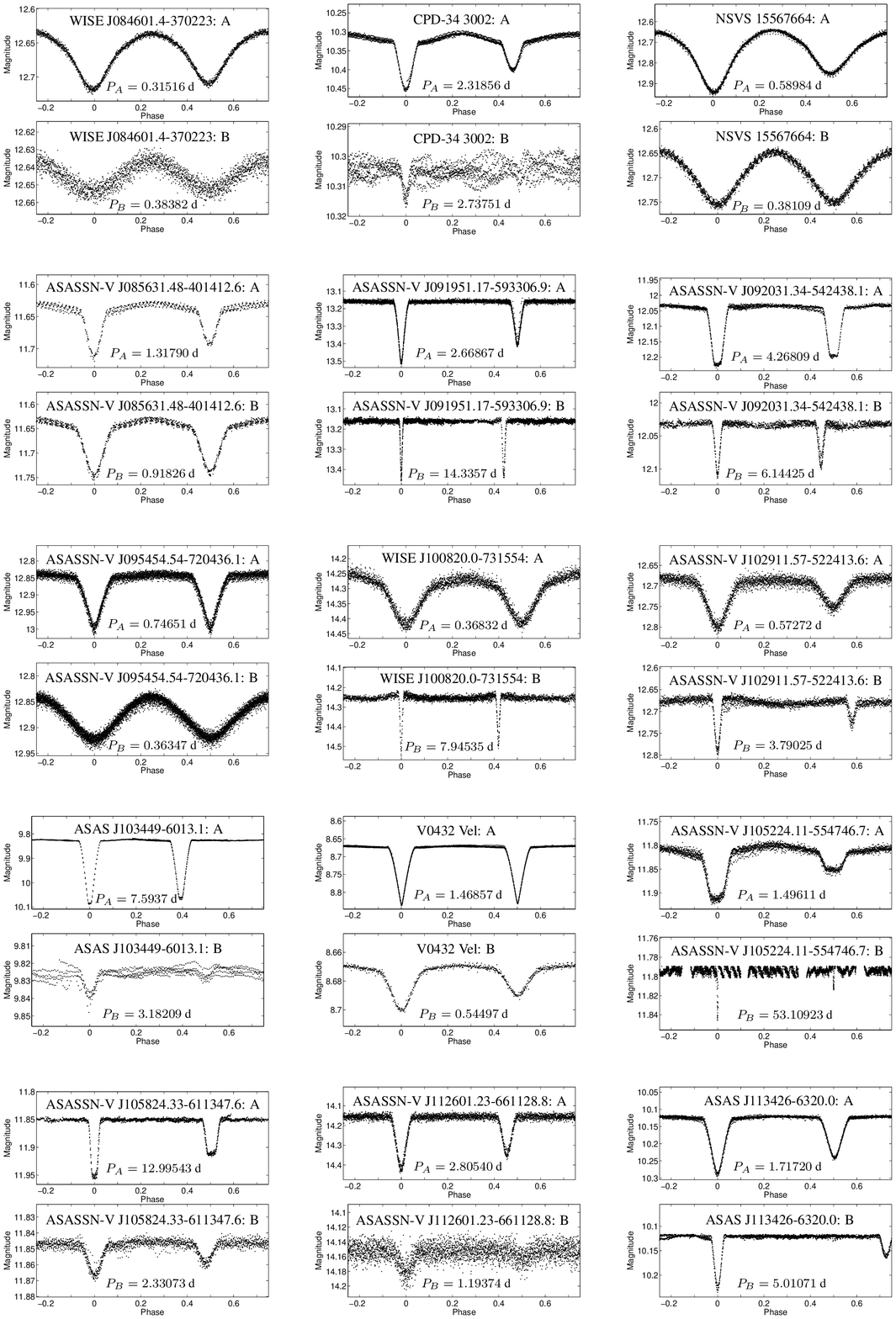}
  \caption{The light curves of both pairs as disentangled from the original TESS photometry, continuation.}
  \label{LC04}
 \end{figure*}

 \begin{figure*}
  \centering
  \includegraphics[width=0.90\textwidth]{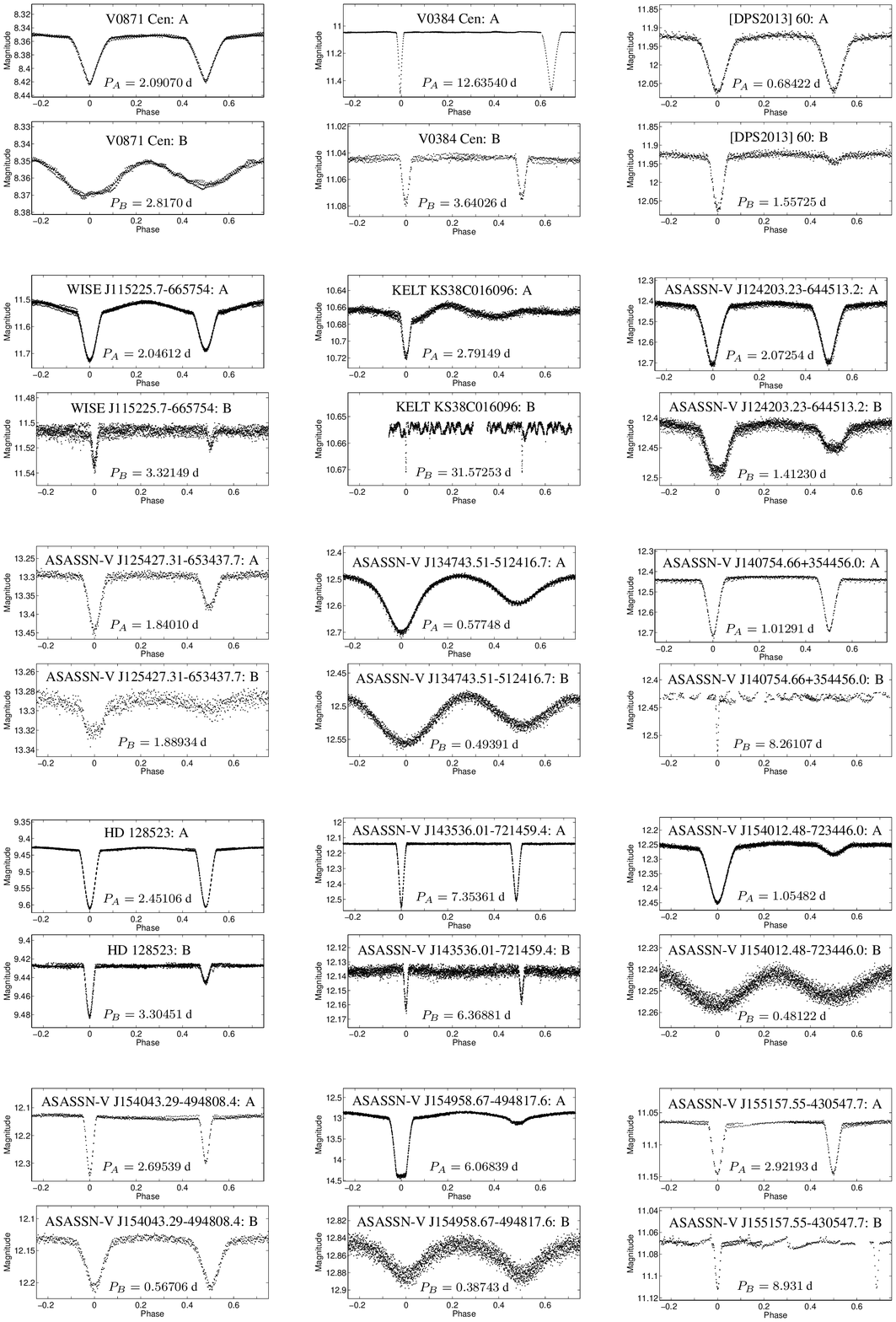}
  \caption{The light curves of both pairs as disentangled from the original TESS photometry, continuation.}
  \label{LC05}
 \end{figure*}

 \begin{figure*}
  \centering
  \includegraphics[width=0.90\textwidth]{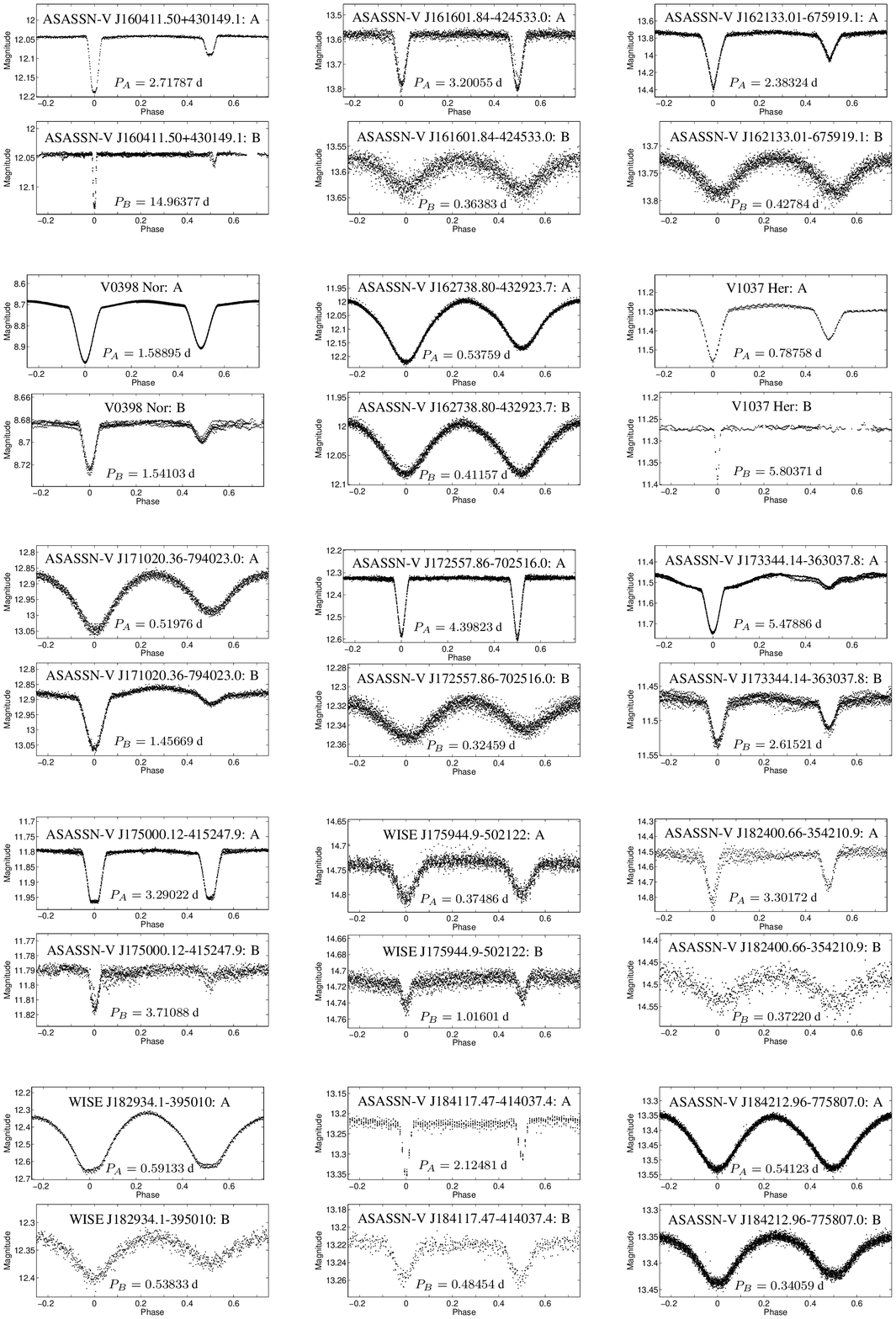}
  \caption{The light curves of both pairs as disentangled from the original TESS photometry, continuation.}
  \label{LC06}
 \end{figure*}

  \begin{figure*}
  \centering
  \includegraphics[width=0.90\textwidth]{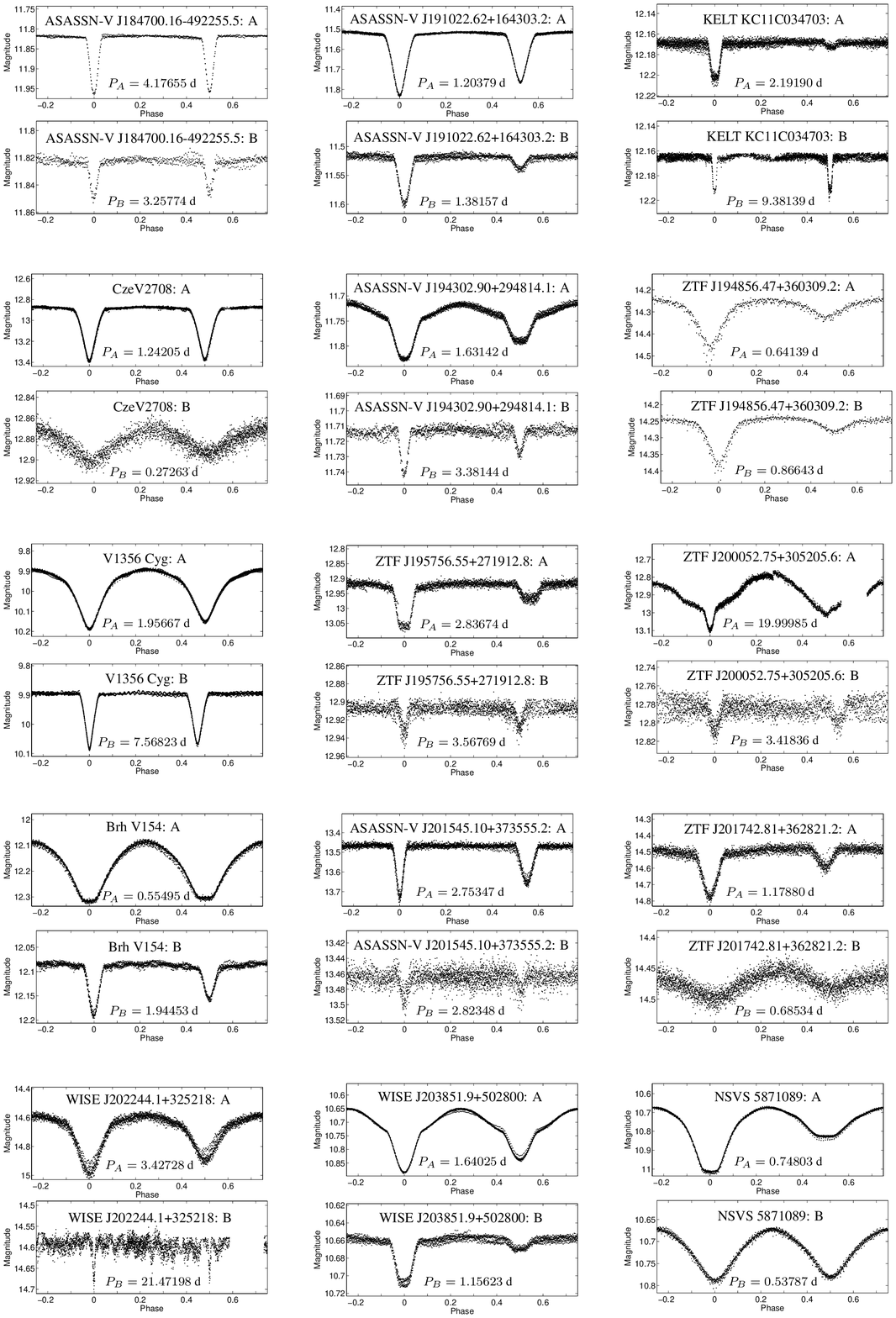}
  \caption{The light curves of both pairs as disentangled from the original TESS photometry, continuation.}
  \label{LC07}
 \end{figure*}

  \begin{figure*}
  \centering
  \includegraphics[width=0.90\textwidth]{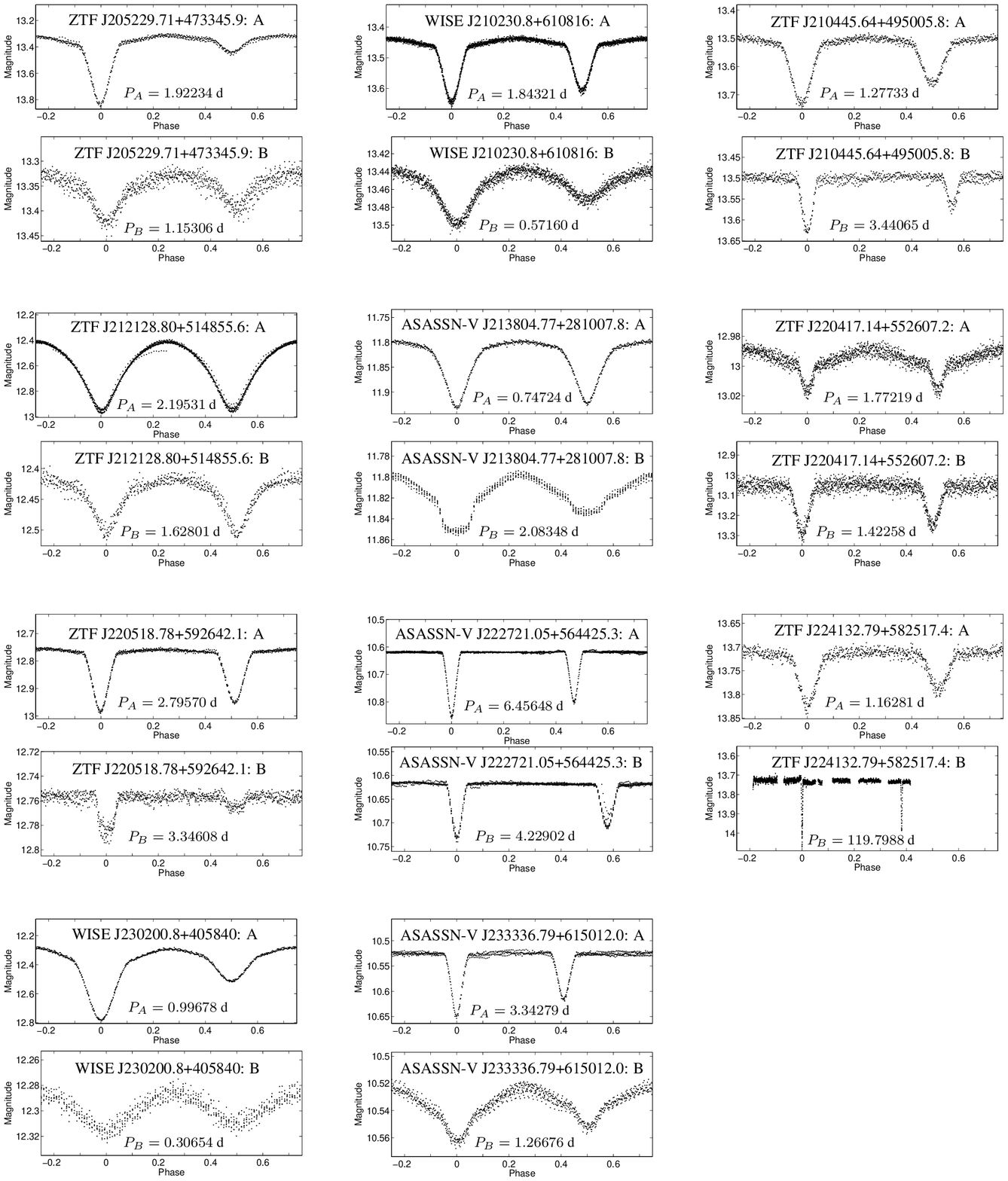}
  \caption{The light curves of both pairs as disentangled from the original TESS photometry, continuation.}
  \label{LC08}
 \end{figure*}

  \begin{figure*}
  \centering
  \includegraphics[width=0.90\textwidth]{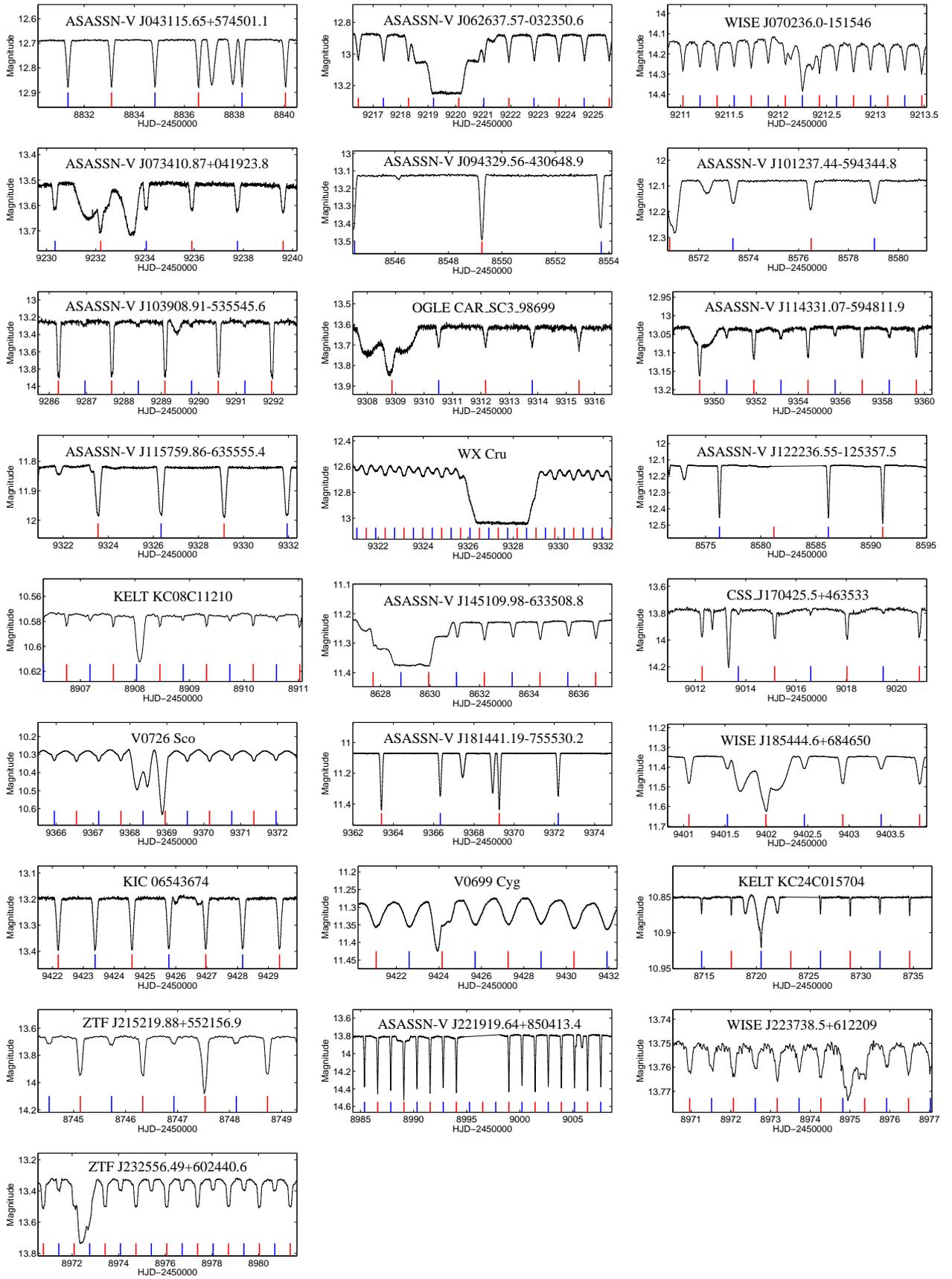}
  \caption{The light curves of triply eclipsing triples. The eclipses of the inner pair are being denoted by short
  abscissae of blue colour for primary, and red for secondary eclipses. Extra eclipses are clearly visible.}
  \label{LCtroj}
 \end{figure*}

 \end{appendix}


\begin{thebibliography}{99}

 \bibitem[Borkovits et al.(2021)]{2021MNRAS.503.3759B} Borkovits, T., Rappaport, S.~A., Maxted, P.~F.~L., et al.\ 2021, \mnras, 503, 3759
 \bibitem[Borkovits et al.(2022)]{2022MNRAS.510.1352B} Borkovits, T., Mitnyan, T., Rappaport, S.~A., et al.\ 2022, \mnras, 510, 1352
 \bibitem[Goodricke(1783)]{1783RSPT...73..474G} Goodricke, J.\ 1783, Philosophical Transactions of the Royal Society of London Series I, 73, 474
 \bibitem[Graczyk et al.(2011)]{2011AcA....61..103G} Graczyk, D., Soszy{\'n}ski, I., Poleski, R., et al.\ 2011, AcA, 61, 103
 \bibitem[Hajdu et al.(2017)]{2017MNRAS.471.1230H} Hajdu, T., Borkovits, T., Forg{\'a}cs-Dajka, E., et al.\ 2017, \mnras, 471, 1230
 \bibitem[Hajdu et al.(2019)]{2019MNRAS.485.2562H} Hajdu, T., Borkovits, T., Forg{\'a}cs-Dajka, E., et al.\ 2019, \mnras, 485, 2562
 \bibitem[Jayaraman et al.(2021)]{2021tsc2.confE..14J} Jayaraman, R., Rappaport, S., Borkovits, T., et al.\ 2021, Posters from the TESS Science Conference II (TSC2), 14
 \bibitem[Kochanek et al.(2017)]{2017PASP..129j4502K} Kochanek, C.~S., Shappee, B.~J., Stanek, K.~Z., et al.\ 2017, \pasp, 129, 104502
 \bibitem[Kostov et al.(2021)]{2021ApJ...917...93K} Kostov, V.~B., Powell, B.~P., Torres, G., et al.\ 2021, \apj, 917, 93
 \bibitem[Kostov et al.(2022)]{2022ApJS..259...66K} Kostov, V.~B., Powell, B.~P., Rappaport, S.~A., et al.\ 2022, \apjs, 259, 66
 \bibitem[Lasker et al.(2008)]{2008AJ....136..735L} Lasker, B.~M., Lattanzi, M.~G., McLean, B.~J., et al.\ 2008, \aj, 136, 735
 \bibitem[Lee et al.(2008)]{2008MNRAS.389.1630L} Lee, C.-U., Kim, S.-L., Lee, J.~W., et al.\ 2008, \mnras, 389, 1630
 \bibitem[Lehmann et al.(2012)]{2012A&A...541A.105L} Lehmann, H., Zechmeister, M., Dreizler, S., et al.\ 2012, \aap, 541, A105
 \bibitem[Lightkurve Collaboration et al.(2018)]{2018ascl.soft12013L} Lightkurve Collaboration, Cardoso, J.~V. de M., Hedges, C., et al.\ 2018, ascl soft, ascl:1812.013
 \bibitem[Masci et al.(2019)]{2019PASP..131a8003M} Masci, F.~J., Laher, R.~R., Rusholme, B., et al.\ 2019, \pasp, 131, 018003
 \bibitem[Masuda et al.(2015)]{2015ApJ...806L..37M} Masuda, K., Uehara, S., \& Kawahara, H.\ 2015, \apjl, 806, L37
 \bibitem[Mayer et al.(1992)]{1992IBVS.3765....1M} Mayer, P., Lorenz, R., \& Drechsel, H.\ 1992, Information Bulletin on Variable Stars, 3765, 1
 \bibitem[Otero(2007)]{2007OEJV...72....1O} Otero, S.~A.\ 2007, Open European Journal on Variable Stars, 0072, 1
 \bibitem[Pawlak et al.(2013)]{2013AcA....63..323P} Pawlak, M., Graczyk, D., Soszy{\'n}ski, I., et al.\ 2013, AcA, 63, 323
 \bibitem[Pollacco et al.(2006)]{2006PASP..118.1407P} Pollacco, D.~L., Skillen, I., Collier Cameron, A., et al.\ 2006, \pasp, 118, 1407
 \bibitem[Rappaport et al.(2022)]{Arxiv2022} Rappaport, S., Borkovits, T., Gagliano, R., et al., 2022, MNRAS, accepted
 \bibitem[Ricker et al.(2015)]{2015JATIS...1a4003R} Ricker, G.~R., Winn, J.~N., Vanderspek, R., et al.\ 2015, JATIS, 1, 014003
 \bibitem[Rozyczka et al.(2011)]{2011MNRAS.414.2479R} Rozyczka, M., Kaluzny, J., Pych, W., et al.\ 2011, \mnras, 414, 2479
 \bibitem[Sana et al.(2011)]{2011MNRAS.416..817S} Sana, H., James, G., \& Gosset, E.\ 2011, \mnras, 416, 817
 \bibitem[Shappee et al.(2014)]{2014ApJ...788...48S} Shappee, B.~J., Prieto, J.~L., Grupe, D., et al.\ 2014, \apj, 788, 48
 \bibitem[Skarka et al.(2017)]{2017OEJV..185....1S} Skarka, M., Ma{\v{s}}ek, M., Br{\'a}t, L., et al.\ 2017, OEJV, 185, 1
 \bibitem[Soszy{\'n}ski et al.(2016)]{2016AcA....66..405S} Soszy{\'n}ski, I., Pawlak, M., Pietrukowicz, P., et al.\ 2016, AcA, 66, 405
 \bibitem[Southworth(2012)]{2012ocpd.conf...51S} Southworth, J.\ 2012, Orbital Couples: Pas de Deux in the Solar System and the Milky Way, 51
 \bibitem[Southworth(2022)]{2022arXiv220102516S} Southworth, J.\ 2022, arXiv:2201.02516
 \bibitem[Tokovinin(2021)]{2021Univ....7..352T} Tokovinin, A.\ 2021, Universe, 7, 352
 \bibitem[Torres et al.(2017)]{2017ApJ...846..115T} Torres, G., Sandberg Lacy, C.~H., Fekel, F.~C., et al.\ 2017, \apj, 846, 115
 \bibitem[Tremaine(2020)]{2020MNRAS.493.5583T} Tremaine, S.\ 2020, \mnras, 493, 5583
 \bibitem[Udalski et al.(1992)]{1992AcA....42..253U} Udalski, A., Szymanski, M., Kaluzny, J., et al.\ 1992, \actaa, 42, 253
 \bibitem[Watson et al.(2006)]{2006SASS...25...47W} Watson, C.~L., Henden, A.~A., \& Price, A.\ 2006, SASS, 25, 47
 \bibitem[Zacharias et al.(2013)]{2013AJ....145...44Z} Zacharias, N., Finch, C.~T., Girard, T.~M., et al.\ 2013, \aj, 145, 44
 \bibitem[Zasche et al.(2009)]{2009AJ....138..664Z} Zasche, P., Wolf, M., Hartkopf, W.~I., et al.\ 2009, \aj, 138, 664
 \bibitem[Zasche et al.(2019)]{2019A&A...630A.128Z} Zasche, P., Vokrouhlick{\'y}, D., Wolf, M., et al.\ 2019, \aap, 630, A128

 \end{thebibliography}
\end{document}